# Age of Information under Source-Aware Truncated ARQ in Multi-Source Wireless Status Updating

Tianci Zhang, Aobo Liu, Zhengchuan Chen, *Senior Member, IEEE,* Zhong Tian, *Member, IEEE,* Jemin Lee, *Senior Member, IEEE,* Yuquan Xiao, and Mehul Motani, *Fellow, IEEE*

***Abstract*—This paper studies information timeliness in multi-source wireless Internet of Things (IoT) status updating systems under a truncated Automatic Repeat reQuest (ARQ) protocol. We propose a source-aware truncated ARQ (SATARQ) scheme that allows differentiated maximum transmission times (MTTs) tailored to different sources. This work focuses on a wireless system with preemptive update management. To study the statistical characteristics of the age of information (AoI) process for each source, a multi-dimensional age process (MDAP) is developed and modeled as a Markov chain, tracking both the AoI and the age of the concerned source's update currently in transmission. Via Markov analysis of the MDAP, we obtain analytical expressions for the distributions and averages of the AoI and peak AoI, as well as the average power consumption of IoT device. The timeliness-energy tradeoff is analyzed by examining the impact of the MTT, update generation probability (UGP), and wireless transmission power (TP). Moreover, this work explores the energy efficiency of the wireless status updating process and its relationship with the information timeliness and energy cost. Numerical results validate the theoretical analysis. Finally, it is demonstrated that the proposed SATARQ, combined with the optimization of MTTs, UGPs, and TPs, significantly improves the overall timeliness-energy tradeoff and energy efficiency across all sources.**

## I. Introduction

### *A. Backgrounds and Related Works*

Real-time wireless status updating typically plays a key role in the time-critical Internet of Things (IoT) applications [2]–[4]. In such wireless systems, the IoT devices (IoTDs) sense the time-sensitive status information of targets, and deliver status updates to edge monitors in real-time, for timely semantics exactions, data analyses, computations, and decision-makings. For instance, in the Internet of Vehicles, the smart vehicles can deliver kinds of vehicular status information such as the velocity and driving environments, via wireless channels to the cooperative vehicles and edge servers, supporting intelligent driver assistant services, vehicle controls, etc. [5]. The quality of the time-critical services typically depends on accuracy and timeliness of the information held at the monitors [3]–[5]. Moreover, in addition to improving the information timeliness, reducing the energy costs (ECs) of IoTDs is another crucial task in the wireless status updating, especially for the energy-limited IoTDs such as remote environmental observation devices. The EC reduction can benefit the sustainability of the wireless status updating and overall system operation.

The age of information (AoI) is a key metric to characterize the information timeliness [6]. The peak AoI (PAoI) is induced from the AoI and utilized for assessing the timeliness in the worst cases [7]. Based on queue modeling, the averages of AoIs and PAoIs in the G/Geo/1 and Ber/G/1 first-come-first-served (FCFS), G/G/1/1 preemptive, and G/G/$\infty$ systems were analyzed [8], [9]. While the averages provide useful measures for timeliness, the AoI and PAoI distributions characterize it completely and rigorously, empowering more comprehensive timeliness studies [10]. For instance, based on the distributions of AoI and PAoI, their violation probabilities [11]–[15] and the statistical AoI [16]–[18] can be conveniently investigated. These AoI-based indicators can stringently evaluate the timeliness for risk-sensitive applications. The distributions of AoI and PAoI can also facilitate the study on the timeliness of the stalest stream among multiple status updating streams [19]. Also, based on the distribution of AoI, its non-linear penalties, e.g., localization errors of mobile IoTDs and remote estimation errors of stochastic processes, can be validly assessed [20]–[23]. The study on AoI and PAoI distributions has attracted a lot of attention [24]–[26]. Based on the multi-dimensional age process (MDAP) method, some works investigated the AoI distributions in the D/G/1/1 preemptive or non-preemptive, Ber/Geo/1 FCFS, Ber/G/1/1 non-preemptive or probabilistically preemptive, and Ber/Geo/1/2 FCFS, preemptive-in-buffer, or probabilistically-preemptive-in-buffer queues [27]–[32]; as well as analyzed the AoI and PAoI distributions in the zero-wait/Geo/1, G/Geo/1/1 preemptive, and multi-source Ber/Geo/1/1 preemptive or non-preemptive systems [33]–[36].

Due to the channel uncertainty in wireless status updating, packet decoding may fail. To improve the information timeliness, it is crucial to manage the updates whose transmissions failed. The retransmission scheme, raising reliability, holds the potential for timeliness enhancement. It provides the monitor the opportunity to successfully receive the failed latest update,

Tianci Zhang is with the School of Microelectronics and Communication Engineering, Chongqing University, Chongqing, China, and also with the Department of Electrical and Computer Engineering, National University of Singapore, Singapore (e-mail: ztc@cqu.edu.cn).

Aobo Liu, Zhengchuan Chen, and Zhong Tian are with the School of Microelectronics and Communication Engineering, Chongqing University, Chongqing, China (e-mails: aobo_liu@stu.cqu.edu.cn, czc@cqu.edu.cn, ztian@cqu.edu.cn).

Jemin Lee is with the School of Electrical and Electronic Engineering, Yonsei University, Seoul, South Korea (e-mail: jemin.lee@yonsei.ac.kr).

Yuquan Xiao is with the School of Information and Communications Engineering, Xi'an Jiaotong University, Shaanxi, China (e-mail: yqxiao@stu.xjtu.edu.cn).

Mehul Motani is with the Department of Electrical and Computer Engineering, National University of Singapore, Singapore (e-mail: motani@nus.edu.sg).

before new updates are generated. Typically, the classical ARQ (CARQ) scheme can be employed, where the IoTD retransmits each update until its correct reception. The CARQ reduces the average AoI and PAoI under the random updating, cooperating with the update preemption [35], [37]. However, this can lead to increased EC for the IoTD due to unlimited retransmissions, which can become excessive under poor channel conditions.

To balance the information timeliness and the EC of IoTD, the truncated ARQ (TARQ) scheme with a maximum transmission time (MTT) can be a promising retransmission strategy and has been employed in wireless status updating [38]–[51].

There have been many works focusing on the information timeliness under the TARQ in the single-source status updating. The timeliness-energy (T-E) tradeoff under the TARQ in the zero-wait system was first analyzed in [38]. Works [39] and [40] studied the averages of AoI, PAoI, and IoTD power under the TARQ in the random-updating preemptive system, as well as optimized the MTT and transmission power (TP) to improve the T-E tradeoff. It was indicated that under the same constraints on average power, the TARQ can enable a better timeliness than the CARQ [39]. The average AoIs and IoTD powers under the TARQ in the random-updating preemptive and non-preemptive systems were compared in [41]. It was found that the update preemption can reduce the AoI. Work [42] proposed to further enhance the T-E tradeoff under the TARQ with the help of receiver diversity. The average AoIs under the TARQ and the truncated hybrid ARQ (THARQ) in the random-updating non-preemptive system with propagation delay were studied in [43]. Work [44] compared the average AoIs and IoTD powers under the TARQ, THARQ, CARQ, and non-ARQ (NARQ) in the zero-wait system. It was indicated that the truncation strategy can achieve a better T-E tradeoff. Considering the delays of coding, transmission, propagation, decoding, and feedback, the AoI and PAoI under the reactive and proactive TARQs and THARQs were analyzed in [45]. Moreover, the TARQ has also been investigated for the relay-aided two-hop status updating systems. It was found that in the zero-wait system, under a given sum MTT, increasing the difference between the MTTs of two hops can reduce the averege AoI, but incur a higher IoTD EC [46]. Works [47] and [48] studied the zero-wait systems where the IoTD is adopted with the NARQ and the direct link is available or not, as well as indicated that the TARQ employment at the relay can improve the timeliness in both scenarios. In addition, the average AoIs and PAoIs in the random-updating non-preemptive relaying systems with or without energy harvesting nodes, where the TARQ or CARQ is used in all hops, were analyzed in [49] and [50]. It was found that the TARQ scheme cooperated with the non-preemptive policy reduces the AoI in the case of frequent update generation [50]. The multi-relay zero-wait system was considered in [51], where the IoTD is adopted with the CARQ and the relays are employed with the TARQ or CARQ. The authors presented TARQ's superiority in achieving the T-E tradeoff for the multi-relay status updating.

### B. Motivations, Novelties, and Challenges

As aforementioned, it is relevant to study the information timeliness under the TARQ for the wireless IoT status updating. In many applications, an IoTD can be a complex unit comprising multiple sensors, resulting in the status updating of multi-source type [52], [53]. In Internet of Vehicles, a vehicle is equipped with kinds of sensors for observing diverse vehicular status data, e.g., the velocity, acceleration, and driving environments. In-depth investigations on the AoI and PAoI statistical characteristics under the TARQ are necessary for the multi-source status updating. *However, the AoI and PAoI under the TARQ have not been studied for the multi-source system.* Moreover, the statistical characteristics of the AoI and PAoI have not been well analyzed under the TARQ, even in the single-source case. *The distributions of the AoI and PAoI under the TARQ have not been reported.* Motivated by these considerations, this work proposes a source-aware TARQ (SATARQ) scheme for the multi-source wireless status updating, and investigates the AoI and PAoI statistical characteristics in a multi-source preemptive system. Under SATARQ, the MTTs can be tailored to the individual sources. The distributions and averages of AoI and PAoI under the SATARQ are derived, and further analysis of the T-E tradeoff and energy efficiency (EE) are carried out. Notably, *i)* the AoI and PAoI as well as the T-E tradeoff and EE under the TARQ are analyzed in multi-source status updating for the first time; and *ii)* the distributions of AoI and PAoI are first characterized for the TARQ scheme.

In addition to the update managements and transmission truncations, the interactions among sources can further affect the AoI behaviors. Compared with the single-source system, the AoI evolution in multi-source system is more complicated, and the statistical characteristics of the AoI and PAoI are more challenging to analyse. To this end, we employ the MDAP method. For each source, the MDAP consisting of the AoI and age of the update in transmission is introduced. It is a discrete-time Markov chain (DTMC). The AoI evolution can be fully tracked via the MDAP, and the statistical characteristics can be studied based on Markov analysis on the MDAP.

### C. Contributions and Organization

The main contributions of this work are:

- The SATARQ scheme with heterogeneous MTTs is introduced for the multi-source wireless IoT status updating. We focus on the typical system with Bernoulli update generation and preemptive update management. This work establishes and investigates the analytical results for probability mass functions (PMFs) and averages of the AoI and PAoI under the SATARQ. Based on the results under SATARQ, the AoI and PAoI results under CARQ and NARQ are also analyzed.
- The average power of the IoTD is assessed. The T-E tradeoff is studied in terms of the MTT, update generation probability (UGP), and TP. It is found that for each source, the variation trends of its timeliness and EC are opposite w.r.t. the MTT and UGP. The opposite variation trends also typically hold w.r.t. the TP. Moreover, the EE is analyzed. The relationship among the EE, timeliness, and EC is deduced. The EE can be regarded as a measure for the T-E tradeoff, and the non-negligible limitations of this operation are demonstrated.
- Numerical results validate the effectiveness of the theoretical analysis. Moreover, it is found that in terms of information

timeliness, the SATARQ can outperform the NARQ remarkably, and can achieve a similar performance as CARQ, even with small MTTs; the SATARQ as well as the optimizations on MTTs, UGPs, and TPs are valid in improving the overall T-E tradeoff and overall EE for all sources. In a shown case, compared with the optimal source-agnostic TARQ, NARQ, and CARQ, the optimal SATARQ can improve the overall T-E tradeoff by about $8.5\%$, $10.3\%$, and $10.3\%$, respectively.

Compared to the preliminary version of this work [1], this paper extends the work in [1] from the following four aspects: *i)* Beyond AoI, this paper studies the statistical characteristics of PAoI to provide a more complete view of the information timeliness. *ii)* In addition to the MTTs, this work further investigates the T-E tradeoff by considering the impacts of the UGPs and TPs. *iii)* This paper additionally studies the EE as well as its relationship with the information timeliness and EC. as well as *iv)* This paper conducts more comprehensive analyses and more detailed demonstrations.

The organization of the rest of this paper is as follows. Section II introduces the system model. The MDAP is analyzed in Section III. Then, Section IV studies the AoI, T-E tradeoff, and EE under the SATARQ. The numerical results and analysis are shown in Section V. At last, Section VI concludes this work.

## II. System Model

Consider a multi-source wireless IoT status updating system, which consists of an IoTD and an edge monitor. The IoTD comprises $N$ sensors and one common transmitter [52]–[54]. The sensors independently sense the status information of $N$ targets and generate status updates. The transmitter delivers the status updates to the edge monitor via a wireless channel. In practical, such IoTD can be a physically integral complicated unit with multiple sensors and one transmitter, e.g., a vehicle [52] or a robot [53], or an abstract integral of multiple sensors and a connected aggregator with one transmitter [54].

The slotted time model is considered [55], where continuous time is divided into consecutive slots. The lengths of all slots are equal to the duration for one time of packet transmission.

At the end of each slot, sensor $i$ (i.e., source $i$) independently generates an update with UGP $q_i$, where $0 \le q_i \le 1$. Let $\boldsymbol{q} := (q_1, \ldots, q_N)$. Each packet transmission starts at the beginning of a slot and completes right before its end. Multiple sources can generate updates at the same slot. The IoTD transmitter can randomly select one of the newly generated updates with the equal probabilities, and identify it as legitimate to transmit [35], [56]. At each slot, the probability that an update of source $i$ is generated and selected to transmit is given by

$$p_i := \sum_{h=0}^{N-1} \sum_{\mathcal{H}\subseteq\mathcal{N}\backslash\{i\},|\mathcal{H}|=h} \frac{1}{h+1} q_i \prod_{j\in\mathcal{H}} q_j \prod_{l\in\mathcal{N}\backslash(\mathcal{H}\bigcup\{i\})} (1-q_l)$$

$$= \frac{q_i(1-p)}{1-q_i} \sum_{h=0}^{N-1} \frac{1}{h+1} \sum_{\mathcal{H}\subseteq\mathcal{N}\backslash\{i\},|\mathcal{H}|=h} \prod_{j\in\mathcal{H}} \frac{q_j}{1-q_j}. \quad (1)$$

Herein, $p$ represents the overall UGP, which is given by

$$p := 1 - \prod_{i=1}^{N} (1-q_i). \quad (2)$$

By definition, it has that $p = \sum_{i=1}^{N} p_i$. The scheme of equal selection probabilities for newly generated updates is only adopted as an example [35], [56]. In effect, the heterogeneous update selection probabilities can also be employed for newly generated updates. Compared to the systems with other probabilistic selection schemes, the only difference lay in $p_i$. All the analysis in this work is also available for these systems.

Because of the channel uncertainty resulted from the fading, noise, interference, etc., the packet transmissions of the status updates could probably fail. Let $\gamma_i$ denote the packet transmission success probability (PTSP) of the updates of source $i$, where $0 \le \gamma_i \le 1$. Note that the data sizes of the updates from different sources can be different. The PTSPs can differ for sources. Due to the channel uncertainty, the SATARQ scheme is employed for enhancing the reliability of transmission and avoiding the excessive EC of the IoTD. Specifically, the update MTTs can be tailored to individual sources. Denote the update MTT for source $i$ by $L_i$. Let $\boldsymbol{L} := (L_1, \ldots, L_N)$. If a packet transmission fails and the elapsed transmission time of current update being transmitted has not reached its MTT, this update will be (further) retransmitted. In contrast, the update whose elapsed transmission time is reaching its MTT will be dropped.

Furthermore, the typical preemptive update management is also employed at the IoTD. Specifically, each newly generated legitimate update starts its transmission at the next slot right after its generation, no matter if there is another update in the service under SATARQ. Notably, the dropping of the non-selected updates and preempted updates can avoid the queueing delay and potentially improve the information timeliness.

To characterize the information timeliness, the AoI is employed. The source-specific AoI is defined as the time elapsed since generation of the latest successfully transmitted update of concerned source. The AoI process of source $i$ is given by

$$\Delta_i(t) := t - U_i(t), \quad (3)$$

where $U_i(t)$ denotes the generation slot of the latest successfully transmitted update of source $i$. The minimum age value is two slots, under the considered timing model [22], [35]. An evolution example of the AoI process can be found in Fig. 1, where all basic system dynamics are included. The PMF of AoI $\Pr\{\Delta_i = n\}$ and average AoI are utilized to measure the information timeliness. The average AoI is given by

$$\overline{\Delta_i} := \mathbb{E}[\Delta_i] = \sum_{n=2}^{+\infty} n \Pr\{\Delta_i = n\}. \quad (4)$$

To assess the timeliness in worst cases, we adopt the PAoI $\mathcal{A}_i$. It represents the AoI when achieving a peak. Let $\{v_i(t) = 1\}$ and $\{v_i(t) = 0\}$ denote that an update of source $i$ is successfully transmitted right before the end of slot $t$ and the opposite. The PMF and average of PAoI of source $i$ can be given by

$$\Pr\{\mathcal{A}_i = n\} := \Pr\{\Delta_i(t) = n | v_i(t) = 1\},\ n \ge 2, \quad (5)$$

$$\overline{\mathcal{A}_i} := \mathbb{E}[\mathcal{A}_i] = \sum_{n=2}^{\infty} n \Pr\{\mathcal{A}_i = n\}. \quad (6)$$

Since the focus of this work is the information timeliness, the performance metrics for characterizing the IoTD EC and EE will be introduced in details when conducting the corresponding analysis, cf. Section IV-B and Section IV-C.

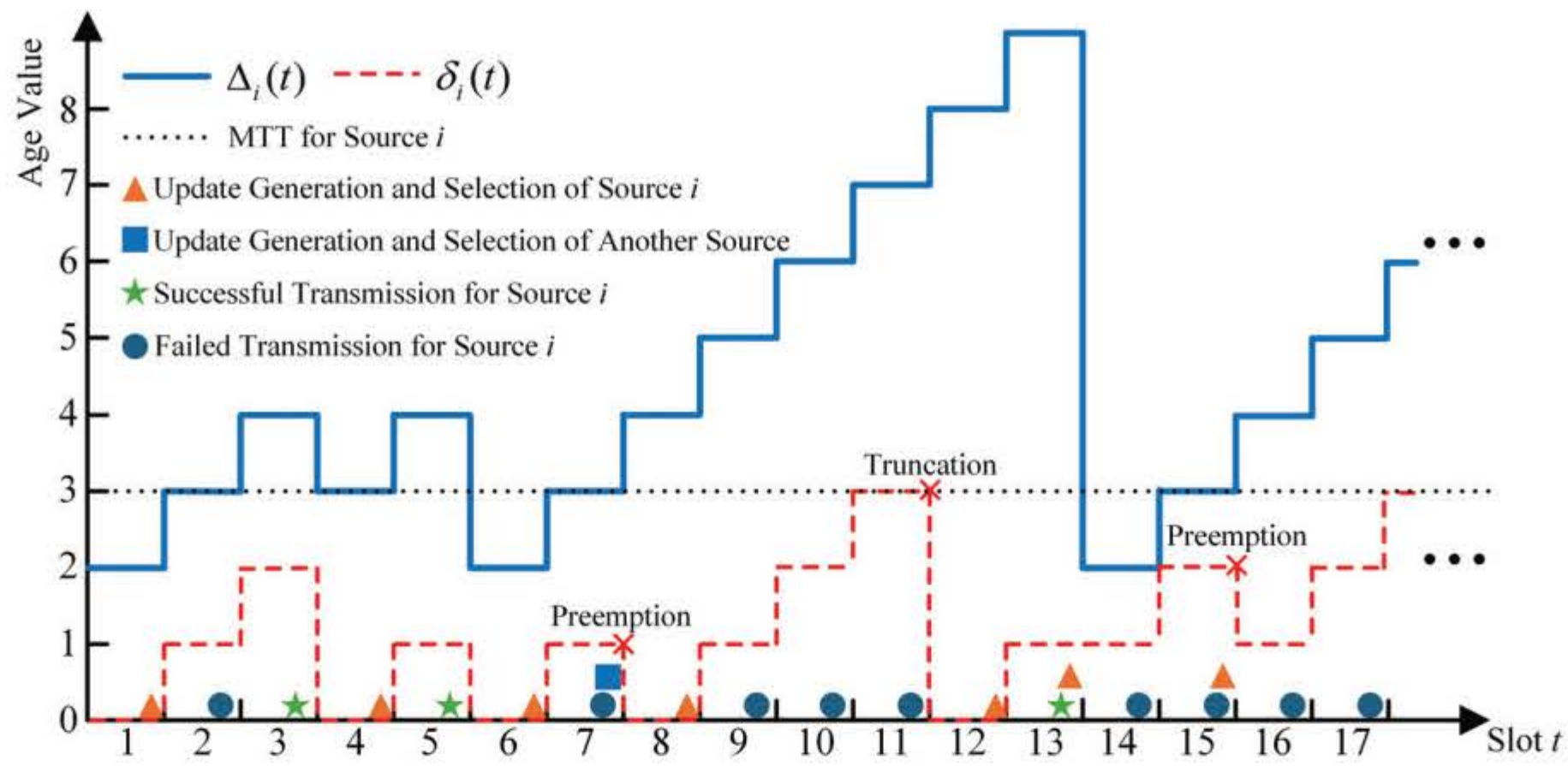

Fig. 1. Evolution examples of the AoI and MDAP. The considered MTT of source $i$ is three slots, i.e., $L_i = 3$.

## III. Markov Analysis on Multi-Dimensional Age Process

This section studies a MDAP for each source, which is a DTMC, consisting of the AoI and the age of the update in transmission. First of all, we introduce and formally define the MDAP. Then, the state transitions of the MDAP are analyzed. Accordingly, its balance equations are formulated. At last, the stationary distribution of the MDAP is obtained. This is the foundation to study the statistical characteristics of AoI and PAoI under the SATARQ as well as the T-E tradeoff and EE.

### A. Definition

Notably, the AoI process under SATARQ in the considered multi-source system is not only affected by the update managements and transmission truncations, but the interactions among sources. Moreover, the AoI does not possess Markov property. The evolution of AoI is complicated, and it is challenging to directly analyse the statistical characteristics of AoI and PAoI. Accordingly, to fully track the evolution of AoI and utilize the underlying Markov property of regarded system, we introduce a MDAP for each source, where the AoI process is one of its elements. Specifically, the MDAP of source $i$ is defined as

$$\mathbf{\Delta}_i(t) := (\Delta_i(t), \delta_i(t)). \tag{7}$$

The supplementary process $\delta_i(t)$ denotes the age of the update of source $i$ in transmission (if any), which also equals to this update's elapsed transmission time right before the end of slot $t$. If no update of source $i$ is being transmitted, let $\delta_i(t) := 0$. Due to the SATARQ scheme, the value space of $\delta_i(t)$ is given by $\{0, 1, \ldots, L_i\}$. Besides, by definition, it has $\Delta_i(t) > \delta_i(t)$. The state space of $\mathbf{\Delta}_i(t)$ is $\{(n, m) : n \geq 2, 0 \leq m \leq L_i, n > m\}$. Notably, one can find that the MDAP is a DTMC.

In the following MDAP analysis, we focus on the common case of $L_i \geq 2$. In fact, the obtained stationary distribution of MDAP also holds for the case of $L_i = 1$. The MDAP analysis in the case of $L_i = 1$, which is omitted here, is similar to and much simpler than that in the case of $L_i \geq 2$.

An evolution example of the MDAP can be found in Fig. 1, where all basic system dynamics are involved.

### B. State Transition

There are three cases of system status: *i)* no update of source $i$ is in transmission at current slot $\tau$ (Case 1); *ii)* an update of source $i$ is being transmitted and its elapsed transmission time has not reached the MTT $L_i$ (Case 2); *iii)* an update of source $i$ is in transmission and its elapsed transmission time is reaching the MTT (Case 3). The state transition in Case 1 depends on if an update of source $i$ is generated and selected at the end of current slot. Those in Case 2 and Case 3 further rely on if the current update transmission of source $i$ succeeds at the current slot, and if an update of another source is generated and selected. This work analyses the state transitions in these three cases respectively. To clarify the state transitions, besides $v_i(t)$, we introduce another indicator process $u_i(t)$. Specifically, let $\{u_i(t) = 1\}$, $\{u_i(t) = 0\}$, and $\{u_i(t) = -1\}$ respectively represent the events that an update of source $i$ is generated and selected at the end of slot $t$, no update is generated, and an update of another source is generated and selected.

*1) State Transitions in Case 1:* In this case, the current state of MDAP is assumed as $\mathbf{\Delta}_i(\tau) = (n-1, 0)$, $n \geq 3$. If $u_i(\tau) = 1$, which happens with probability $p_i$, the newly generated update of source $i$ will be transmitted at the next slot. This results in $\mathbf{\Delta}_i(\tau+1) = (n, 1)$. If $u_i(\tau) = 0$ or $u_i(\tau) = -1$, which happens with probability $(1 - p_i)$, there is still no update of source $i$ being transmitted at the next slot. It has that $\mathbf{\Delta}_i(\tau+1) = (n, 0)$.

*2) State Transitions in Case 2:* In this case, one can assume that $\mathbf{\Delta}_i(\tau) = (n-1, m-1)$, $n > m$, $2 \leq m \leq L_i$. If $v_i(\tau) = 1$ and $u_i(\tau) = 1$, the status information from source $i$ maintained at the monitor will be refreshed, and the new legitimate update of source $i$ will be transmitted at the next slot. Accordingly, $\mathbf{\Delta}_i(\tau+1) = (m, 1)$. This happens with probability $\gamma_i p_i$. If $v_i(\tau) = 1$ and $u_i(\tau) = 0$ or $-1$, it has $\mathbf{\Delta}_i(\tau+1) = (m, 0)$. This happens with probability $\gamma_i(1 - p_i)$. If $v_i(\tau) = 0$ and $u_i(\tau) = 1$, which happens with probability $(1 - \gamma_i)p_i$, the current update is preempted and the IoTD will transmit the new legitimate update of source $i$ instead. This results in $\mathbf{\Delta}_i(\tau+1) = (n, 1)$. If $v_i(\tau) = 0$ and $u_i(\tau) = -1$, the current update of source $i$ is preempted by the new legitimate update of another source. Accordingly, $\mathbf{\Delta}_i(\tau+1)=(n,0)$. This happens with probability $(1-\gamma_i)(p-p_i)$. If $v_i(\tau)=0$ and $u_i(\tau)=0$, the current update will be retransmitted. It has $\mathbf{\Delta}_i(\tau+1)=(n,m)$.

TABLE I
STATE TRANSITIONS AND TRANSITION PROBABILITIES OF MDAP

| Current State | $u_i(\tau)$ | $v_i(\tau)$ | Next State | Transition Probability | Ranges of $n$ and $m$ |
|---|---|---|---|---|---|
| $(n-1,0)$ | 1 | 0 | $(n,1)$ | $p_i$ | $n\geq 3$ |
| $(n-1,0)$ | 0 or $-1$ | 0 | $(n,0)$ | $1-p_i$ | $n\geq 3$ |
| $(n-1,m-1)$ | 1 | 1 | $(m,1)$ | $\gamma_i p_i$ | $2\leq m\leq L_i+1,\ n>m$ |
| $(n-1,m-1)$ | 0 or $-1$ | 1 | $(m,0)$ | $\gamma_i(1-p_i)$ | $2\leq m\leq L_i+1,\ n>m$ |
| $(n-1,m-1)$ | 1 | 0 | $(n,1)$ | $(1-\gamma_i)p_i$ | $2\leq m\leq L_i+1,\ n>m$ |
| $(n-1,m-1)$ | $-1$ | 0 | $(n,0)$ | $(1-\gamma_i)(p-p_i)$ | $2\leq m\leq L_i,\ n>m$ |
| $(n-1,m-1)$ | 0 | 0 | $(n,m)$ | $(1-\gamma_i)(1-p)$ | $2\leq m\leq L_i,\ n>m$ |
| $(n-1,L_i)$ | 0 or $-1$ | 0 | $(n,0)$ | $(1-\gamma_i)(1-p_i)$ | $n\geq L_i+2$ |

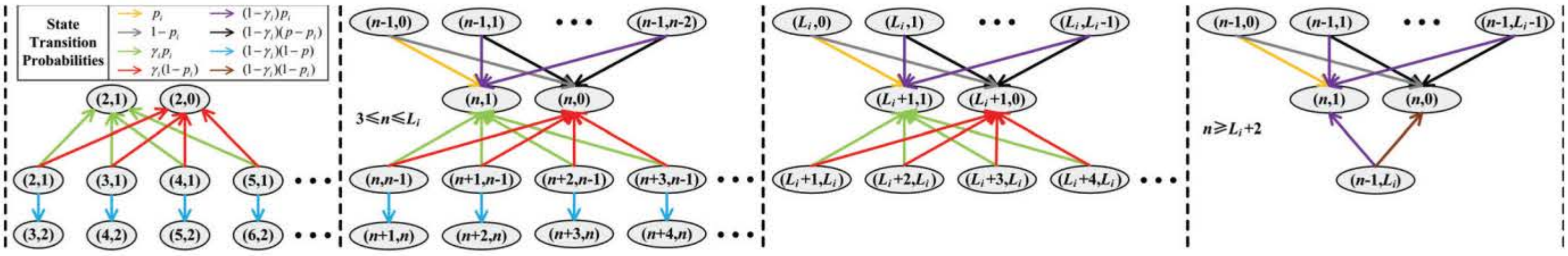

Fig. 2. State transition diagram of MDAP where $L_i \geq 3$. By deleting the second part of diagram, the state transition diagram where $L_i = 2$ can be obtained.

*3) State Transitions in Case 3:* The current state of MDAP can be assumed as $\mathbf{\Delta}_i(\tau)=(n-1,L_i)$, $n\geq L_i+2$. If $v_i(\tau)=1$, the system operations and the MDAP state transitions in Case 3 can be similar to those in Case 2. Specifically, *i)* if $v_i(\tau)=1$ and $u_i(\tau)=1$, $\mathbf{\Delta}_i(\tau+1)=(L_i+1,1)$; as well as *ii)* if $v_i(\tau)=1$ and $u_i(\tau)=0$ or $-1$, $\mathbf{\Delta}_i(\tau+1)=(L_i+1,0)$. These happen with probabilities $\gamma_i p_i$ and $\gamma_i(1-p_i)$ respectively. If $v_i(\tau)=0$, the SATARQ scheme takes effect, the transmission of current update must be truncated, and no retransmission will be further provided for it. Accordingly, *i)* if $v_i(\tau)=0$ and $u_i(\tau)=1$, $\mathbf{\Delta}_i(\tau+1)=(n,1)$; as well as *ii)* if $v_i(\tau)=0$ and $u_i(\tau)=0$ or $-1$, $\mathbf{\Delta}_i(\tau+1)=(n,0)$. The state transition probabilities are given by $(1-\gamma_i)p_i$ and $(1-\gamma_i)(1-p_i)$ respectively.

For more clarity, the state transitions and transition probabilities of the MDAP are completely summarized in Table I. Fig. 2 presents the state transition diagram of the MDAP.

### *C. Balance Equations and Stationary Distribution*

Let $\pi_s^{(i)}$ denote the stationary distribution of MDAP, where $s$ is the concerned state. According to Table I and Fig. 2, the balance equations of the MDAP can be formulated as (8a)–(8g), which are presented at the top of the next page.

By solving the balance equations as well as utilizing the regularization condition of probability distribution, the stationary distribution of MDAP can be obtained, cf. Proposition 1.

**Proposition 1:** The stationary distribution of the MDAP under the SATARQ can by completely given by

$$\pi_{(n,m)}^{(i)} = \lambda_i^{m-1} y_{n-m+1}^{(i)},\ n>m,\ 1\leq m\leq L_i, \tag{9}$$

$$\pi_{(n,0)}^{(i)} = g_n^{(i)},\ n\geq 2, \tag{10}$$

$$Y^{(i)}(z) = \frac{\gamma_i p_i^2 (1-(\lambda_i/z)^{L_i})}{(z-1)(z-\lambda_i)+\gamma_i p_i z(1-(\lambda_i/z)^{L_i})}, \tag{11}$$

$$G^{(i)}(z) = \frac{\gamma_i p_i (1-(\lambda_i/z)^{L_i})(z-\lambda_i-p_i z(1-(\lambda_i/z)^{L_i}))}{(z-\lambda_i)((z-1)(z-\lambda_i)+\gamma_i p_i z(1-(\lambda_i/z)^{L_i}))}. \tag{12}$$

Herein, $\lambda_i = (1-\gamma_i)(1-p)$; $y_n^{(i)} = g_n^{(i)} = 0$, $n\leq 1$; $Y^{(i)}(z)$ and $G^{(i)}(z)$ denote the $z$-transforms of $y_n^{(i)}$ and $g_n^{(i)}$, respectively.

*Proof:* First of all, by repeatedly iterating (8g) and using $\pi_{(n,1)}^{(i)} = \lambda_i^{1-1}\pi_{(n-1+1,1)}^{(i)}$, it can be deduced that

$$\pi_{(n,m)}^{(i)} = \lambda_i^{m-1}\pi_{(n-m+1,1)}^{(i)},\ n>m,\ 1\leq m\leq L_i. \tag{13}$$

Accordingly, to obtain the complete stationary distribution of MDAP, the core is to find $\pi_{(n,1)}^{(i)}$ and $\pi_{(n,0)}^{(i)}$. Let us introduce auxiliary sequences $y_n^{(i)}$ and $g_n^{(i)}$, step sequence $u_n$, as well as window sequence $w_n^{(i)}$. They are respectively defined as

$$y_n^{(i)} := \begin{cases} \pi_{(n,1)}^{(i)}, & n\geq 2 \\ 0, & n\leq 1 \end{cases}, \tag{14}$$

$$g_n^{(i)} := \begin{cases} \pi_{(n,0)}^{(i)}, & n\geq 2 \\ 0, & n\leq 1 \end{cases}, \tag{15}$$

$$u_n := \begin{cases} 1, & n\geq 0 \\ 0, & n\leq -1 \end{cases}, \tag{16}$$

$$w_n^{(i)} := \begin{cases} 1, & 0\leq n\leq L_i-1 \\ 0, & n\leq -1 \text{ or } n\geq L_i \end{cases}. \tag{17}$$

It has $w_n^{(i)} = u_n - u_{n-L_i}$. It can be found that acquiring $\pi_{(n,1)}^{(i)}$ and $\pi_{(n,0)}^{(i)}$ is equivalent to deriving $y_n^{(i)}$ and $g_n^{(i)}$. Furthermore, sequences $y_n^{(i)}$ and $g_n^{(i)}$ can be identified by their $z$-transforms.

Next, let us focus on the derivations on $z$-transforms of $y_n^{(i)}$ and $g_n^{(i)}$, i.e., $Y^{(i)}(z)$ and $G^{(i)}(z)$. By plugging (8a) and (13) into (8c)–(8f), as well as utilizing (14)–(17), one can obtain

$$y_n^{(i)} = p_i g_{n-1}^{(i)} + \lambda_i^{n-2}\pi_{(2,1)}^{(i)} + (1-\gamma_i)p_i \sum_{k=1}^{n-2} \lambda_i^{k-1} y_{n-k}$$

$$\pi_{(2,1)}^{(i)} = \gamma_i p_i \sum_{n=2}^{+\infty} \pi_{(n,1)}^{(i)}, \tag{8a}$$

$$\pi_{(2,0)}^{(i)} = \gamma_i (1-p_i) \sum_{n=2}^{+\infty} \pi_{(n,1)}^{(i)}, \tag{8b}$$

$$\pi_{(n,1)}^{(i)} = p_i \pi_{(n-1,0)}^{(i)} + \gamma_i p_i \sum_{l=n+1}^{+\infty} \pi_{(l-1,n-1)}^{(i)} + (1-\gamma_i) p_i \sum_{m=2}^{n-1} \pi_{(n-1,m-1)}^{(i)}, \ 3 \le n \le L_i + 1, \tag{8c}$$

$$\pi_{(n,1)}^{(i)} = p_i \pi_{(n-1,0)}^{(i)} + (1-\gamma_i) p_i \sum_{m=2}^{L_i+1} \pi_{(n-1,m-1)}^{(i)}, \ n \ge L_i + 2, \tag{8d}$$

$$\pi_{(n,0)}^{(i)} = (1-p_i) \pi_{(n-1,0)}^{(i)} + \gamma_i (1-p_i) \sum_{l=n+1}^{+\infty} \pi_{(l-1,n-1)}^{(i)} + (1-\gamma_i)(p-p_i) \sum_{m=2}^{n-1} \pi_{(n-1,m-1)}^{(i)}, \ 3 \le n \le L_i + 1, \tag{8e}$$

$$\pi_{(n,0)}^{(i)} = (1-p_i) \pi_{(n-1,0)}^{(i)} + (1-\gamma_i)(1-p) \pi_{(n-1,L_i)}^{(i)} + (1-\gamma_i)(p-p_i) \sum_{m=2}^{L_i+1} \pi_{(n-1,m-1)}^{(i)}, \ n \ge L_i + 2, \tag{8f}$$

$$\pi_{(n,m)}^{(i)} = (1-\gamma_i)(1-p) \pi_{(n-1,m-1)}^{(i)}, \ 2 \le m \le L_i, \ n > m. \tag{8g}$$

---

$$\begin{aligned} &= p_i g_{n-1}^{(i)} + \lambda_i^{n-2} \pi_{(2,1)}^{(i)} + (1-\gamma_i) p_i \\ &\quad \times (\lambda_i^{n-1} u_{n-1}) * y_n^{(i)}, \ 3 \le n \le L_i + 1, \end{aligned} \tag{18}$$

$$\begin{aligned} y_n^{(i)} &= p_i g_{n-1}^{(i)} + (1-\gamma_i) p_i \sum_{k=1}^{L_i} \lambda_i^{k-1} y_{n-k} \\ &= p_i g_{n-1}^{(i)} + (1-\gamma_i) p_i \\ &\quad \times (\lambda_i^{n-1} w_{n-1}^{(i)}) * y_n^{(i)}, \ n \ge L_i + 2, \end{aligned} \tag{19}$$

$$\begin{aligned} g_n^{(i)} &= (1-p_i) g_{n-1}^{(i)} + \frac{1-p_i}{p_i} \lambda_i^{n-2} \pi_{(2,1)}^{(i)} + (1-\gamma_i)(p-p_i) \\ &\quad \times (\lambda_i^{n-1} u_{n-1}) * y_n^{(i)}, \ 3 \le n \le L_i + 1, \end{aligned} \tag{20}$$

$$\begin{aligned} g_n^{(i)} &= (1-p_i) g_{n-1}^{(i)} + \lambda_i^{L_i} y_{n-L_i}^{(i)} + (1-\gamma_i)(p-p_i) \\ &\quad \times (\lambda_i^{n-1} w_{n-1}^{(i)}) * y_n^{(i)}, \ n \ge L_i + 2, \end{aligned} \tag{21}$$

where $*$ denotes the convolution operator w.r.t. $n$. In effect, it is found that: *i)* $y_n^{(i)} = 0$ for $n \le 1$; *ii)* $g_n^{(i)} = 0$ for $n \le 1$; *iii)* $(\lambda_i^{n-1} u_{n-1}) * y_n^{(i)} = 0$ for $n \le 2$; *iv)* $(\lambda_i^{n-1} u_{n-L_i-1}) * y_n^{(i)} = 0$ for $n \le L_i + 2$, i.e., $(\lambda_i^{n-1} w_{n-1}^{(i)}) * y_n^{(i)} = (\lambda_i^{n-1} u_{n-1}) * y_n^{(i)}$ for $n \le L_i + 2$; and *v)* $y_{n-L_i}^{(i)} = 0$ for $n \le L_i + 1$. Denote the set of all integers by $\mathbb{Z}$. Combining these critical findings and (18)–(21), the general expressions of $y_n^{(i)}$ and $g_n^{(i)}$, which hold for all integer values of $n$, can be respectively given by

$$\begin{aligned} y_n^{(i)} &= p_i g_{n-1}^{(i)} + \lambda_i^{n-2} w_{n-2}^{(i)} \pi_{(2,1)}^{(i)} + (1-\gamma_i) p_i \\ &\quad \times (\lambda_i^{n-1} w_{n-1}^{(i)}) * y_n^{(i)}, \ n \in \mathbb{Z}, \end{aligned} \tag{22}$$

$$\begin{aligned} g_n^{(i)} &= (1-p_i) g_{n-1}^{(i)} + \lambda_i^{L_i} y_{n-L_i}^{(i)} + \frac{1-p_i}{p_i} \lambda_i^{n-2} w_{n-2}^{(i)} \pi_{(2,1)}^{(i)} \\ &\quad + (1-\gamma_i)(p-p_i)(\lambda_i^{n-1} w_{n-1}^{(i)}) * y_n^{(i)}, \ n \in \mathbb{Z}. \end{aligned} \tag{23}$$

Accordingly, by performing $z$-transform on the both sides of (22) and (23), the equations w.r.t. the $z$-transforms of $y_n^{(i)}$ and $g_n^{(i)}$, i.e., $Y^{(i)}(z)$ and $G^{(i)}(z)$, can be formulated as

$$\begin{aligned} Y^{(i)}(z) &= p_i z^{-1} G^{(i)}(z) + \frac{1-(\lambda_i/z)^{L_i}}{1-(\lambda_i/z)} \\ &\quad \times (z^{-2} \pi_{(2,1)}^{(i)} + (1-\gamma_i) p_i z^{-1} Y^{(i)}(z)), \end{aligned} \tag{24a}$$

$$\begin{aligned} G^{(i)}(z) &= (1-p_i) z^{-1} G^{(i)}(z) + (\lambda_i/z)^{L_i} Y^{(i)}(z) \\ &\quad + \frac{1-(\lambda_i/z)^{L_i}}{1-\lambda_i/z} \Big( \frac{1-p_i}{p_i} z^{-2} \pi_{(2,1)}^{(i)} \\ &\quad + (1-\gamma_i)(p-p_i) z^{-1} Y^{(i)}(z) \Big). \end{aligned} \tag{24b}$$

By solving equations (24a)–(24b), $Y^{(i)}(z)$ and $G^{(i)}(z)$ can be respectively determined by

$$Y^{(i)}(z) = \frac{(1-(\lambda_i/z)^{L_i}) \pi_{(2,1)}^{(i)}}{(z-1)(z-\lambda_i) + \gamma_i p_i z (1-(\lambda_i/z)^{L_i})}, \tag{25}$$

$$G^{(i)}(z) = \Big( \frac{1}{p_i} - \frac{1-(\lambda_i/z)^{L_i}}{1-\lambda_i/z} \Big) Y^{(i)}(z). \tag{26}$$

It remains to find $\pi_{(2,1)}^{(i)}$. Based on (13) and the regularization condition of stationary distribution, it has that

$$\begin{aligned} 1 &= \sum_{n=2}^{+\infty} \pi_{(n,0)}^{(i)} + \sum_{n=2}^{+\infty} \sum_{m=1}^{L_i} \pi_{(n,m)}^{(i)} \\ &= \sum_{n=2}^{+\infty} \pi_{(n,0)}^{(i)} + \sum_{n=2}^{+\infty} (\lambda_i^n w_n^{(i)}) * y_n \\ &= G^{(i)}(1) + \frac{1-(\lambda_i/z)^{L_i}}{1-\lambda_i/z} \Big|_{z=1} Y^{(i)}(1). \end{aligned} \tag{27}$$

Based on (25) and (26), it can be deduced from (27) that

$$\pi_{(2,1)}^{(i)} = \gamma_i p_i^2 \tag{28}$$

By combining (25), (26), and (28), the $z$-transforms of auxiliary sequences $y_n^{(i)}$ and $g_n^{(i)}$ can be derived, cf. (11) and (12).

Finally, based on (13)–(15), the complete stationary distribution can be given by (9)–(12). This concludes the proof. ■

## IV. Age of Information, Timeliness-Energy Tradeoff, and Energy Efficiency

This section investigates the statistical characteristics of AoI and PAoI, T-E tradeoff, as well as the EE under the SATARQ.

First, based on the stationary distribution of the MDAP, the PMFs and averages of the AoI and PAoI are obtained. The average power of the IoTD is also found via the MDAP. Then, the T-E tradeoff under the SATARQ is analyzed in terms of the MTT, UGP, and TP. Finally, the EE and its relationship with the information timeliness and EC are studied.

### *A. Statistical Characteristics of Age of Information and Peak Age of Information*

*1) PMF and Average of AoI:* Since the AoI process is the first element of MDAP, the PMF of AoI can be expressed as

$$\begin{aligned}\Pr\{\Delta_i = n\} =& \Pr\left\{\bigcup_{m=0}^{n-1}\{\mathbf{\Delta}_i = (n,m)\}\right\}\\ =&\pi^{(i)}_{(n,0)} + \sum_{m=1}^{n-1}\pi^{(i)}_{(n,m)},\ 2\le n\le L_i+1, \quad (29)\\ \Pr\{\Delta_i = n\} =& \Pr\left\{\bigcup_{m=0}^{L_i}\{\mathbf{\Delta}_i = (n,m)\}\right\}\\ =&\pi^{(i)}_{(n,0)} + \sum_{m=1}^{L_i}\pi^{(i)}_{(n,m)},\ n\ge L_i+2. \quad (30)\end{aligned}$$

According to (29), (30) and Proposition 1, the AoI statistical characteristics can be obtained, as presented in Theorem 1.

**Theorem 1:** The PMF and average of the AoI under the SATARQ can be respectively given by

$$\Pr\{\Delta_i = n\} =\phi^{(i)}_n,\ n\ge 2, \quad (31)$$

$$\Phi^{(i)}(z) =\frac{\gamma_i p_i\big(1-(\lambda_i/z)^{L_i}\big)}{(z-1)(z-\lambda_i)+\gamma_i p_i z(1-(\lambda_i/z)^{L_i})}, \quad (32)$$

$$\overline{\Delta_i} =\frac{1-\lambda_i}{\gamma_i p_i(1-\lambda_i{}^{L_i})}+1, \quad (33)$$

where $\phi^{(i)}_n = 0,\ n\le 1$, and $\Phi^{(i)}(z)$ is the $z$-transform of $\phi^{(i)}_n$.

*Proof:* First, let us derive the PMF of AoI. We introduce an auxiliary sequence $\phi^{(i)}_n$, which is defined as

$$\phi^{(i)}_n := \begin{cases}\Pr\{\Delta_i = n\}, & n\ge 2\\ 0, & n\le 1\end{cases}. \quad (34)$$

Accordingly, acquiring the PMF of AoI can be equivalent to deriving $\phi^{(i)}_n$, which is identified by its $z$-transform $\Phi^{(i)}(z)$.

In the following, let us derive the $z$-transform of $\phi^{(i)}_n$. By substituting (9) into (29) and (30), it can be obtained that

$$\begin{aligned}\phi^{(i)}_n &= g^{(i)}_n + \sum_{k=0}^{n-2}\lambda_i^k y_{n-k}\\ &= g^{(i)}_n + (\lambda_i^n u_n) * y^{(i)}_n,\ 2\le n\le L_i+1, \quad (35)\\ \phi^{(i)}_n &= g^{(i)}_n + \sum_{k=0}^{L_i-1}\lambda_i^k y_{n-k}\\ &= g^{(i)}_n + (\lambda_i^n w^{(i)}_n) * y^{(i)}_n,\ n\ge L_i+2. \quad (36)\end{aligned}$$

It is notable that *i)* $\phi^{(i)}_n = 0$ for $n\le 1$; *ii)* $g^{(i)}_n = 0$ for $n\le 1$; *iii)* $(\lambda_i^n u_n) * y^{(i)}_n = 0$ for $n\le 1$; and *iv)* $(\lambda_i^n u_{n-L_i}) * y^{(i)}_n = 0$ for $n\le L_i+1$, i.e., $(\lambda_i^n w_n) * y^{(i)}_n = (\lambda_i^n u_n) * y^{(i)}_n$ for $n\le L_i+1$. Combining these with (35) and (36), the general expression of $\phi^{(i)}_n$, holding for all integer values of $n$, can be given by

$$\phi^{(i)}_n = g^{(i)}_n + (\lambda_i^n w^{(i)}_n) * y^{(i)}_n,\ n\in\mathbb{Z}. \quad (37)$$

By performing $z$-transform on both sides of (37), it has that

$$\Phi^{(i)}(z) = G^{(i)}(z) + \frac{1-(\lambda_i/z)^{L_i}}{1-\lambda_i/z}Y^{(i)}(z). \quad (38)$$

By substituting (11) and (12) into (38), $\Phi^{(i)}(z)$ and the PMF of AoI can be obtained, as given by (31) and (32).

Moreover, the average AoI can be determined by

$$\overline{\Delta_i} = \sum_{n=-\infty}^{+\infty} n\phi^{(i)}_n = -\frac{\mathrm{d}\Phi^{(i)}(z)}{\mathrm{d}z}\Big|_{z=1}. \quad (39)$$

Based on (32) and (39), the average AoI is given by (33). ■

*2) PMF and Average of PAoI:* In effect, the PAoI PMF can be determined based on the PMF of AoI and the stationary distribution of MDAP. Specifically, when an update of source $i$ is successfully transmitted, there must be an update of source $i$ is in transmission at the current slot. That is, $\{v_i(t)=1\}\subset\{\delta_i(t)\ge 1\}$. This deduces $\{v_i(t)=1\}=\{\delta_i(t)\ge 1\}\cap\{v_i(t)=1\}$. Thus, based on (5), the PMF of PAoI can be expressed as

$$\begin{aligned}\Pr\{\mathcal{A}_i = n\} =& \Pr\{\Delta_i(t) = n|\delta_i(t)\ge 1, v_i(t)=1\}\\ =&\frac{\Pr\{\Delta_i(t)=n,\delta_i(t)\ge 1\}}{\Pr\{\delta_i(t)\ge 1\}}\\ &\times\frac{\Pr\{v_i(t)=1|\Delta_i(t)=n,\delta_i(t)\ge 1\}}{\Pr\{v_i(t)=1|\delta_i(t)\ge 1\}}\\ =&\frac{\Pr\{\Delta_i=n\}-\pi^{(i)}_{(n,0)}}{1-\sum_{n=2}^{+\infty}\pi^{(i)}_{(n,0)}},\ n\ge 2. \quad (40)\end{aligned}$$

The last equality holds following from $\Pr\{v_i(t)=1|\Delta_i(t)=n,\delta_i(t)\ge 1\} = \Pr\{v_i(t)=1|\delta_i(t)\ge 1\} = \gamma_i$. Based on (40), Proposition 1, and Theorem 1, the PMF and average of the PAoI can be obtained, as shown in Theorem 2.

**Theorem 2:** The PMF and average of the PAoI under the SATARQ can be respectively given by

$$\Pr\{\mathcal{A}_i=n\}=\psi^{(i)}_n,\ n\ge 2, \quad (41)$$

$$\Psi^{(i)}(z)=\frac{\gamma_i p_i\big(\frac{1-\lambda_i}{1-\lambda_i^{L_i}}\big)z(1-(\lambda_i/z)^{L_i})^2}{(z-\lambda_i)((z-1)(z-\lambda_i)+\gamma_i p_i z(1-(\lambda_i/z)^{L_i}))}, \quad (42)$$

$$\overline{\mathcal{A}_i}=\frac{1-\lambda_i}{\gamma_i p_i(1-\lambda_i^{L_i})}+\frac{1}{1-\lambda_i}-\frac{L_i\lambda_i^{L_i}}{1-\lambda_i^{L_i}}, \quad (43)$$

where $\psi^{(i)}_n = 0,\ n\le 1$, and $\Psi^{(i)}(z)$ is the $z$-transform of $\psi^{(i)}_n$.

*Proof:* Similar to acquiring the PMF of AoI, to obtain the PMF of PAoI, let us utilize a proper auxiliary sequence and derive its $z$-transform. The auxiliary sequence is defined as

$$\psi^{(i)}_n := \begin{cases}\Pr\{\mathcal{A}_i = n\}, & n\ge 2\\ 0, & n\le 1\end{cases}. \quad (44)$$

Note that $\psi^{(i)}_n = \psi^{(i)}_n = g^{(i)}_n = 0$ for $n\le 1$. Eq. (40) indicates

$$\psi^{(i)}_n = \frac{\phi^{(i)}_n - g^{(i)}_n}{1-\sum_{n=-\infty}^{+\infty} g^{(i)}_n},\ n\in\mathbb{Z}. \quad (45)$$

Accordingly, the $z$-transform of $\psi_n^{(i)}$ can be determined by

$$\Psi^{(i)}(z) = \frac{\Phi^{(i)}(z) - G^{(i)}(z)}{1 - G^{(i)}(1)}. \quad (46)$$

By plugging (12) and (32) into (46), $\Psi^{(i)}(z)$ and the PMF of PAoI can be obtained, as presented in (41) and (42).

Accordingly, the average PAoI can be expressed as

$$\overline{\mathcal{A}_i} = \sum_{n=-\infty}^{+\infty} n\psi_n^{(i)} = -\frac{\mathrm{d}\Psi^{(i)}(z)}{\mathrm{d}z}\bigg|_{z=1}. \quad (47)$$

By substituting (42) into (47), Eq. (43) can be obtained. ■

It can be found that the AoI and PAoI statistical characteristics of any source under the SATARQ depend on the MTT of this source only, and are independent from those of other sources. The independence results from the preemptive update management. This renders a significant flexibility for adjusting the MTTs. One can design a proper MTT for each source and no impact can be incurred on the timeliness of other sources.

Moreover, both the CARQ and NARQ can be regarded as the special cases of the SATARQ. Specifically, *i)* the CARQ can be seen as the SATARQ with the infinite MTTs, i.e., $L_i = +\infty$ for $\forall i \in \{1, \ldots, N\}$; and *ii)* the NARQ can be seen as the SATARQ with the unit MTTs, i.e., $L_i = 1$ for $\forall i \in \{1, \ldots, N\}$. With the help of Theorem 1 and Theorem 2, the statistical characteristics of the AoIs and PAoIs under the CARQ and NARQ can be deduced, as given in the following corollary.

**Corollary 1:** The PMFs and averages of the AoI and PAoI under the CARQ can be respectively given by

$$\Pr\{\Delta_i^{\mathrm{C}} = n\} = \phi_n^{\mathrm{C},(i)},\ n \geq 2,\ \Pr\{\mathcal{A}_i^{\mathrm{C}} = n\} = \psi_n^{\mathrm{C},(i)},\ n \geq 2, \quad (48)$$

$$\Phi^{\mathrm{C},(i)}(z) = \frac{\gamma_i p_i}{z(z - (1 - \gamma_i p_i)) - \lambda_i(z - 1)}, \quad (49)$$

$$\Psi^{\mathrm{C},(i)}(z) = \frac{\gamma_i p_i (1 - \lambda_i) z}{(z - \lambda_i)(z(z - (1 - \gamma_i p_i)) - \lambda_i(z - 1))}, \quad (50)$$

$$\overline{\Delta_i^{\mathrm{C}}} = \frac{1 - \lambda_i}{\gamma_i p_i} + 1,\ \overline{\mathcal{A}_i^{\mathrm{C}}} = \frac{1 - \lambda_i}{\gamma_i p_i} + \frac{1}{1 - \lambda_i}. \quad (51)$$

Those under the NARQ can be respectively given by

$$\Pr\{\Delta_i^{\mathrm{N}} = n\} = \phi_n^{\mathrm{N},(i)},\ n \geq 2,\ \Pr\{\mathcal{A}_i^{\mathrm{N}} = n\} = \psi_n^{\mathrm{N},(i)},\ n \geq 2, \quad (52)$$

$$\Phi^{\mathrm{N},(i)}(z) = \Psi^{\mathrm{N},(i)}(z) = \frac{\gamma_i p_i}{z(z - (1 - \gamma_i p_i))}, \quad (53)$$

$$\overline{\Delta_i^{\mathrm{N}}} = \overline{\mathcal{A}_i^{\mathrm{N}}} = \frac{1}{\gamma_i p_i} + 1. \quad (54)$$

Herein, $\phi_n^{\mathrm{C},(i)} = \phi_n^{\mathrm{N},(i)} = \psi_n^{\mathrm{C},(i)} = \psi_n^{\mathrm{N},(i)} = 0$ for $n \leq 1$; as well as $\Phi^{\mathrm{C},(i)}(z)$, $\Phi^{\mathrm{N},(i)}(z)$, $\Psi^{\mathrm{C},(i)}(z)$, and $\Psi^{\mathrm{N},(i)}(z)$ are the $z$-transforms of $\phi_n^{\mathrm{C},(i)}$, $\phi_n^{\mathrm{N},(i)}$, $\psi_n^{\mathrm{C},(i)}$, and $\psi_n^{\mathrm{N},(i)}$, respectively.

*Proof:* By plugging $L_i = \infty$ or $L_i = 1$ for $\forall i \in \{1, \ldots, N\}$ into (31)–(33) and (41)–(43), Eqs. (49)–(51) or (52)–(54) can be obtained with some mathematical operations. ■

Under the NARQ, the statistical characteristics of AoI and PAoI are completely the same, cf. (52)–(54). It is an interesting finding against the intuition that the PAoI tends to take larger values than AoI's. The reason of this behavior is demonstrated as follows. Let $\{d_i(t) = 1\}$ denote the event that an update of source $i$ is generated and selected at the end of slot $(t-1)$ and it is successfully transmitted right before the end of slot $t$. The complementary event is denoted by $\{d_i(t) = 0\}$. According to the NARQ scheme, the AoI evolution can be given by

$$\Delta_i^{\mathrm{N}}(t) = \begin{cases} 2, & d_i(t-1) = 1 \\ \Delta_i^{\mathrm{N}}(t-1) + 1, & d_i(t-1) = 0 \end{cases}. \quad (55)$$

This reveals that the AoI at slot $t$ is uniquely determined by $\{d_i(\zeta)\}_{\zeta \leq t-1}$. Moreover, under the NARQ, by definition, it has that $d_i(t) = v_i(t)$. The PAoI PMF can be reexpressed as

$$\Pr\{\mathcal{A}^{\mathrm{N},(i)} = n\} = \Pr\{\Delta^{\mathrm{N},(i)}(t) = n | d_i(t) = 1\}. \quad (56)$$

Note that indicator $d_i(t)$ is an independent stochastic process. This reveals that $\Delta^{\mathrm{N},(i)}(t)$ is independent of $d_i(t)$. Combining these results, one can deduce that under the NARQ, the PMFs, i.e., statistical characteristics of the AoI and PAoI are the same.

### *B. Timeliness-Energy Tradeoff*

In many IoT applications, e.g., the real-time environmental monitoring for remote areas, the IoTD is often battery-based and energy-limited. In addition to improving the information timeliness, reducing the EC of IoTD is also significant. However, there is a tradeoff between the timeliness and EC. In the following, let us analyse the T-E tradeoff under the SATARQ.

It is notable that the retransmission strategy is introduced to wireless status updating for improving the information timeliness, and the TARQ is further developed to avoid excessive retransmissions and saving the EC of packet transmissions. Accordingly, we focus on transmission EC, as done in TARQ studies [39], [42], and [49]. In fact, the transmission EC can be typically the main kind of EC for IoTD, as often considered in literatures where even the TARQ is not the focus [57]–[60]. Other kinds of ECs can be deemed relatively negligible [39].

Denote the duty cycle and TP of IoTD for transmitting the status updates of source $i$ by $\rho_i$ and $P_i$, respectively. Let $\boldsymbol{P} := (P_1, \ldots, P_N)$. Note that $\{\delta_i(t) \geq 1\}$ represents that the IoTD is transmitting an update of source $i$ at slot $t$. Based on the MDAP distribution presented in Proposition 1, the source-specific duty cycle under the SATARQ can be given by

$$\begin{aligned} \rho_i &:= \lim_{T \to +\infty} \frac{1}{T} \sum_{t=1}^{T} \mathbb{I}\{\delta_i(t) \geq 1\} \\ &= \Pr\{\delta_i(t) \geq 1\} \\ &= 1 - \sum_{n=2}^{+\infty} \pi_{(n,0)}^{(i)} \\ &= \frac{p_i(1 - \lambda_i{}^{L_i})}{1 - \lambda_i}. \end{aligned} \quad (57)$$

It is noteworthy that $\gamma_i = \gamma_i(P_i)$, $p = p(\boldsymbol{q})$, $p_i = p_i(\boldsymbol{q})$, and $\lambda_i = (1 - p(\boldsymbol{q}))(1 - \gamma_i(P_i))$. Accordingly, the source-specific duty cycle is a function of the MTT and TP of this source and the UGPs of all sources, i.e., $\rho_i = \rho_i(L_i, \boldsymbol{q}, \gamma_i(P_i))$. Therefore, the source-specific average power of IoTD, i.e., the average per-slot energy use to deliver the updates of a certain source, characterizing the source-specific EC, can be given by

$$E_i(L_i, \boldsymbol{q}, P_i) := \lim_{T \to +\infty} \frac{1}{T} \sum_{t=1}^{T} P_i \mathbb{I}\{\delta_i(t) \geq 1\}$$

$$=P_i \rho_i(L_i, \boldsymbol{q}, \gamma_i(P_i)) \tag{58}$$

$$=P_i \frac{p_i(1-{\lambda_i}^{L_i})}{1-\lambda_i}. \tag{59}$$

Also, based on Theorem 1 and (57), the average AoI, which is also a lower bound of average PAoI, can be expressed as

$$\overline{\Delta_i}(L_i, \boldsymbol{q}, P_i) = \frac{1}{\gamma_i(P_i)} \cdot \frac{1}{\rho_i(L_i, \boldsymbol{q}, \gamma_i(P_i))} + 1. \tag{60}$$

Eqs. (58) and (60) explicitly characterize the source-specific T-E tradeoff under the SATARQ scheme, in terms of the MTT $L_i$, UGPs $\boldsymbol{q}$, and TP $P_i$. It is found that the variation trends of the source-specific average AoI $\overline{\Delta_i}$ and average power $E_i$ w.r.t. the MTT and UGPs are always opposite. This is because that $L_i$ and $\boldsymbol{q}$ can affect only $\rho_i$ in $\overline{\Delta_i}$ and $E_i$. In particular, the average AoI is always decreasing with the increase of the MTT, while the average power always rises. The relationship between their variation trends w.r.t. the TP could be more complicated. Specifically, the average AoI $\overline{\Delta_i}$ is decreasing w.r.t. TP $P_i$, since the PTSP $\gamma_i$ is increasing w.r.t. $P_i$ as well as $\overline{\Delta_i}$ is decreasing w.r.t. $\gamma_i$. In contrast, the average power $E_i$ could be non-monotonous w.r.t. TP $P_i$. Nevertheless, in common cases, the average power $E_i$ is increasing w.r.t. TP $P_i$, also possessing the opposite variation trend w.r.t. $P_i$ compared with the average AoI. This fact can be revealed by Proposition 2, where typical models of channel and PTSP are considered.

**Proposition 2:** Consider a Rayleigh block-fading channel contaminated by Gaussian noise of unit variance, where the channel gain $h$ follows the standard complex Gaussian distribution. Let $1-\gamma_i(P_i)$ stand for the channel outage probability. It is found that *i)* in the case of $L_i \le 3$ or $p \ge \frac{1}{e^2+1}$, the source-specific average power $E_i$ is always increasing w.r.t. the TP $P_i$; and *ii)* when $\gamma_i(P_i) \ge \min\{1-\frac{1}{2(1-p)}, \frac{1}{e}\}$, average power $E_i$ is increasing w.r.t. $P_i$. Herein, $\gamma_i(P_i) = e^{-\frac{k_i}{P_i}}$, $k_i := e^{R_i} - 1$, and $R_i$ denotes the channel coding rate for the source $i$'s updates.

*Proof:* First of all, since the power gain $|h|^2$ follows the exponential distribution of unit expectation in the considered channel model, it can be obtained that

$$\gamma_i(P_i) = \Pr\{R_i \le \ln(1 + P_i|h|^2)\} = e^{-\frac{k_i}{P_i}}. \tag{61}$$

Notably, $\lambda_i = (1-p)(1-\gamma_i(P_i))$. The first-order derivative of source-specific average power $E_i$ w.r.t. TP $P_i$ is given by

$$\frac{\mathrm{d}E_i}{\mathrm{d}P_i} = p_i \Big( \frac{1-\lambda_i^{L_i}}{1-\lambda_i} - (1-p)\frac{k_i}{P_i} e^{-\frac{k_i}{P_i}} \times \Big( \frac{-L_i \lambda_i^{L_i-1}}{1-\lambda_i} + \frac{1-\lambda_i^{L_i}}{(1-\lambda_i)^2} \Big) \Big). \tag{62}$$

Note that $xe^{-x} \le 1 - e^{-x}$, which is equivalent to $e^x \ge x+1$. Accordingly, it has that $(1-p)\frac{k_i}{P_i}e^{-\frac{k_i}{P_i}} \le (1-p)(1-e^{-\frac{k_i}{P_i}}) = \lambda_i$. By substituting this inequality into (62), it is obtained that

$$\frac{\mathrm{d}E_i}{\mathrm{d}P_i} \ge \frac{p_i}{(1-\lambda_i)^2}(1 - 2\lambda_i + \lambda_i^{L_i+1} + (L_i - 1)\lambda_i^{L_i}(1-\lambda_i)). \tag{63}$$

In the cases of $L_i = 1$, $L_i = 2$, and $L_i = 3$, Eq. (63) indicates that $\frac{\mathrm{d}E_i}{\mathrm{d}P_i} \ge p_i \ge 0$, $\frac{\mathrm{d}E_i}{\mathrm{d}P_i} \ge p_i \ge 0$, and $\frac{\mathrm{d}E_i}{\mathrm{d}P_i} \ge p_i(1-\lambda_i^2) \ge 0$, respectively. In the case of $L_i \le 3$, the monotonicity of average power $E_i$ w.r.t. $P_i$ is found. Also, Eq. (63) reveals that when $\lambda_i \le \frac{1}{2}$, i.e., $\gamma_i(P_i) \ge 1 - \frac{1}{2(1-p)}$, $E_i$ is increasing w.r.t. TP $P_i$.

Moreover, based on (62) and $-L_i\lambda_i^{L_i-1}(1-\lambda_i) \le 0$, one can obtain another critical inequality of $\frac{\mathrm{d}E_i}{\mathrm{d}P_i}$. It is given by

$$\frac{\mathrm{d}E_i}{\mathrm{d}P_i} \ge \frac{p_i(1-p)(1-\lambda_i^{L_i})}{(1-\lambda_i)^2} \Big( \frac{1}{1-p} - 1 + \Big(1 - \frac{k_i}{P_i}\Big) e^{-\frac{k_i}{P_i}} \Big). \tag{64}$$

Since that $1/(1-p) \ge 1$, Eq. (64) reveals that

$$\frac{\mathrm{d}E_i}{\mathrm{d}P_i} \ge \frac{p_i(1-p)(1-\lambda_i^{L_i})}{(1-\lambda_i)^2} \Big(1 - \frac{k_i}{P_i}\Big) e^{-\frac{k_i}{P_i}}. \tag{65}$$

Accordingly, it has that when $\frac{k_i}{P_i} \le 1$, i.e., $\gamma_i(P_i) \ge \frac{1}{e}$, average power $E_i$ is increasing w.r.t. $P_i$. By using $(1-x)e^{-x} \ge -e^{-2}$, which is equivalent to $e^{x-2} \ge x - 1$, Eq. (64) also reveals that

$$\frac{\mathrm{d}E_i}{\mathrm{d}P_i} \ge \frac{p_i(1+e^{-2})(1-\lambda_i^{L_i})}{(1-\lambda_i)^2} \Big( p - \frac{1}{e^2+1} \Big). \tag{66}$$

This indicates that in the case of $p \ge \frac{1}{e^2+1}$, the average power $E_i$ is always increasing w.r.t. TP $P_i$. This concludes the proof. ■

The source-specific T-E tradeoff analysis also reveals that the MTTs, UGPs, and TPs can be optimized to improve the overall T-E tradeoff of all sources. Typically and effectively, *i)* the overall timeliness can be represented by the source-average average AoI, i.e., the arithmetic average of all sources' average AoIs; and *ii)* the overall EC of IoTD can be evaluated by the average IoTD power. They are respectively given by

$$\overline{\Delta_{\mathrm{S}}}(\boldsymbol{L}, \boldsymbol{q}, \boldsymbol{P}) := \frac{1}{N}\sum_{i=1}^{N} \overline{\Delta_i}$$

$$= \frac{1}{N}\sum_{i=1}^{N} \frac{1}{\gamma_i(P_i)} \cdot \frac{1}{\rho_i(L_i, \boldsymbol{q}, \gamma_i(P_i))} + 1 \tag{67}$$

$$= \frac{1}{N}\sum_{i=1}^{N} \frac{1-\lambda_i}{\gamma_i p_i (1 - {\lambda_i}^{L_i})} + 1, \tag{68}$$

$$E_{\mathrm{O}}(\boldsymbol{L}, \boldsymbol{q}, \boldsymbol{P}) := \lim_{T\to+\infty} \frac{1}{T}\sum_{t=1}^{T}\sum_{i=1}^{N} P_i \mathbb{I}\{\delta_i(t) \ge 1\}$$

$$= \sum_{i=1}^{N} P_i \rho_i(L_i, \boldsymbol{q}, \gamma_i(P_i)) \tag{69}$$

$$= \sum_{i=1}^{N} P_i \frac{p_i(1-{\lambda_i}^{L_i})}{1-\lambda_i}. \tag{70}$$

Typically, this work employs the weighted sum (WS) of these two indicators as a valid characterization for the overall T-E tradeoff. The effectiveness of optimizing the MTTs, UGPs, and TPs for improving the overall T-E tradeoff of all sources will be demonstrated through numerical results in Section V.[1]

[1]This work pays attention to the analytical characterizations for the information timeliness and EC under the SATARQ, as well as the analysis for T-E tradeoff. In terms of the system optimization under the SATARQ, this work demonstrates the effectiveness of optimizing the MTTs, UGPs, and TPs for improving the overall T-E tradeoff, via numerical results. The optimization algorithm design is out of the focus of this work, and can be further studied in-depth in future works with using the results of this work as foundations.

### C. Energy Efficiency

The popular definition of the EE is the average number of successfully delivered bits per energy use. Note that typically, the updating of the status information at monitor rather than its data amount is the focus for status updating tasks. We study the EE representing the average number of successfully delivered updates per energy use. The source-specific EE, i.e., the EE for transmitting updates of a certain source, is defined as

$$\eta_i := \lim_{T\to+\infty} \frac{M_i(T)}{\sum_{t=1}^{T} P_i \mathbb{I}\{\delta_i(t)\geq 1\}}, \tag{71}$$

where stochastic process $M_i(t)$ stands for the number of the successfully transmitted updates of source $i$ until slot $t$. Denote the number of successfully delivered updates of all sources till slot $t$ by $M(t)$. The overall EE for all sources is defined as

$$\eta_{\mathrm{O}} := \lim_{T\to+\infty} \frac{M(T)}{\sum_{t=1}^{T}\sum_{i=1}^{N} P_i \mathbb{I}\{\delta_i(t)\geq 1\}}, \tag{72}$$

In the following, let us analyse the analytical expressions of the source-specific and overall EEs, as well as investigate the relationship among the EE, information timeliness, and EC.

*1) Source-Specific EE:* Let $X_i^{(j)}$ denote the $j$-th interval of generations and selections of consecutive successful updates of source $i$. The source-specific EE can be determined by

$$\eta_i = \frac{\lim_{T\to+\infty}\frac{1}{\frac{\sum_{j=1}^{M_i(T)} X_i^{(j)}}{M_i(T)}}}{\lim_{T\to+\infty}\frac{\sum_{t=1}^{T} P_i \mathbb{I}\{\delta_i(t)\geq 1\}}{T}} = \frac{1}{E_i\mathbb{E}[X_i]}. \tag{73}$$

To acquire the EE, it remains to find the average of $X_i$. In fact, it can be complicated and challenging to directly analyse the statistical characteristics of $X_i$. For concision, this work adopts an indirect approach which utilizes the results of PAoI. Specifically, let $T_i$ denote the update transmission time of a successful update of source $i$. There is an explicit relationship among the averages of PAoI, $X_i$, and $T_i$, as given by

$$\overline{\mathcal{A}_i} = \mathbb{E}[X_i] + \mathbb{E}[T_i]. \tag{74}$$

Note that $\overline{\mathcal{A}_i}$ has been derived. One can obtain the average of $X_i$ through deriving that of $T_i$. Let $S_i^{(l)}$ and $\mathcal{E}_i^{(l)}$ respectively denote the total packet transmission times of the $l$-th generated and selected update of source $i$, and the event that this update is successfully delivered. The PMF of $T_i$ is determined by

$$\begin{aligned}\Pr\{T_i=n\} &= \Pr\{S_i^{(l)}=n|\mathcal{E}_i^{(l)}\}\\ &= \frac{\Pr\{S_i^{(l)}=n,\mathcal{E}_i^{(l)}\}}{\sum_{n=1}^{L_i}\Pr\{S_i^{(l)}=n,\mathcal{E}_i^{(l)}\}},\ 1\leq n\leq L_i. \end{aligned}\tag{75}$$

Herein, the numerator represents the probability that the $l$-th update of source $i$ is transmitted for $n$ times and successfully delivered at the last attempt. Thus, the numerator is given by

$$\begin{aligned}\Pr\{S_i^{(l)}=n,\mathcal{E}_i^{(l)}\} &= (1-p)^{n-1}(1-\gamma_i)^{n-1}\gamma_i\\ &= \lambda_i^{n-1}\gamma_i,\ 1\leq n\leq L_i. \end{aligned}\tag{76}$$

Combining (75)-(76), the PMF and average of $T_i$ are given by

$$\Pr\{T_i=n\} = \frac{1-\lambda_i}{1-\lambda_i^{L_i}}\lambda_i^{n-1},\ 1\leq n\leq L_i. \tag{77}$$

$$\mathbb{E}[T_i] = \frac{1-\lambda_i}{1-\lambda_i^{L_i}}\sum_{n=1}^{L_i} n\lambda_i^{n-1} = \frac{1}{1-\lambda_i} - \frac{L_i\lambda_i^{L_i}}{1-\lambda_i^{L_i}}. \tag{78}$$

Based on (43), (74), and (78), the average of $X_i$ is given by

$$\mathbb{E}[X_i] = \frac{1-\lambda_i}{\gamma_i p_i(1-{\lambda_i}^{L_i})}. \tag{79}$$

By plugging (59) and (79) into (73), the analytical expression of the source-specific EE is finally obtained, as given by

$$\eta_i = \frac{\gamma_i}{P_i}. \tag{80}$$

Notably, $\gamma_i=\gamma_i(P_i)$. The source-specific EE is a function w.r.t. the TP for concerned source, i.e., $\eta_i=\eta_i(P_i)$.

According to Theorem 1 and (59), the relationship among source-specific EE, information timeliness, and EC is deduced. Specifically, the source-specific EE can be expressed as

$$\eta_i = \frac{1}{(\overline{\Delta_i}-1)E_i}. \tag{81}$$

The source-specific EE could be regarded as a kind of characterization for the source-specific T-E tradeoff. Unfortunately, Eq. (80) indicates that different from the source-specific average AoI and average power, the source-specific EE is only determined by the TP $P_i$ and is independent of the MTT $L_i$ and UGPs $\boldsymbol{q}$. The effects of system design on source-specific T-E tradeoff cannot be fully reflected via the source-specific EE. The source-specific EE cannot sufficiently characterize the source-specific tradeoff. This is a non-negligible limitation of source-specific EE for measuring source-specific T-E tradeoff.

*2) Overall EE:* Notably, by definition, $M(t)=\sum_{i=1}^{N} M_i(t)$. Thus, based on (59) and (79), the overall EE can be given by

$$\begin{aligned}\eta_{\mathrm{O}} &= \frac{\sum_{i=1}^{N}\frac{1}{\lim_{T\to+\infty}\frac{\sum_{j=1}^{M_i(T)} X_i^{(j)}}{M_i(T)}}}{\sum_{i=1}^{N}\lim_{T\to+\infty}\frac{\sum_{t=1}^{T} P_i \mathbb{I}\{\delta_i(t)\geq 1\}}{T}}\\ &= \frac{\sum_{i=1}^{N}\frac{1}{\mathbb{E}[X_i]}}{\sum_{i=1}^{N} E_i}\\ &= \frac{\sum_{i=1}^{N}\gamma_i\frac{p_i(1-{\lambda_i}^{L_i})}{1-\lambda_i}}{\sum_{i=1}^{N} P_i\frac{p_i(1-{\lambda_i}^{L_i})}{1-\lambda_i}}. \end{aligned}\tag{82}$$

Note that $\gamma_i=\gamma_i(P_i)$, $p_i=p_i(\boldsymbol{q})$, and $\lambda_i=\lambda_i(\boldsymbol{q},P_i)$. The overall EE is a function of the MTTs, UGPs, and TPs of all sources. That is, $\eta_{\mathrm{O}}=\eta_{\mathrm{O}}(\boldsymbol{L},\boldsymbol{q},\boldsymbol{P})$.

Based on Theorem 1, (70), and (82), the relationship among the overall EE, data timeliness, and EC can be given by

$$\eta_{\mathrm{O}} = \frac{N}{\frac{N}{\sum_{i=1}^{N}\frac{1}{\overline{\Delta_i}-1}}E_{\mathrm{O}}}. \tag{83}$$

In (83), the item $\frac{N}{\sum_{i=1}^{N}\frac{1}{\overline{\Delta_i}-1}}$, which is the harmonic average of $\{\overline{\Delta_i}-1\}_{i=1}^{N}$, could be considered as a kind of measure for the overall timeliness of all sources. Accordingly, similar to the source-specific case, the overall EE could be regarded as a kind of assessment for the overall T-E tradeoff. Although the effects of all of the MTTs, UGPs, and TPs could be reflected via the overall EE, measuring the overall T-E tradeoff through the overall EE still possesses non-negligible limitations. Particularly, when improving the overall T-E tradeoff through maximizing the overall EE, the optimal UGPs of some sources under given MTTs and TPs as well as some optimal TPs under given MTTs and UGPs can be zero. This will be demonstrated in details based on the numerical results in Section V. Any zero UGP or zero TP is unexpected. The monitor can acquire no update of the sources with zero UGPs or zero TPs, resulting in the devastation on the information timeliness of these sources and even the possible operation breakdown of the system.

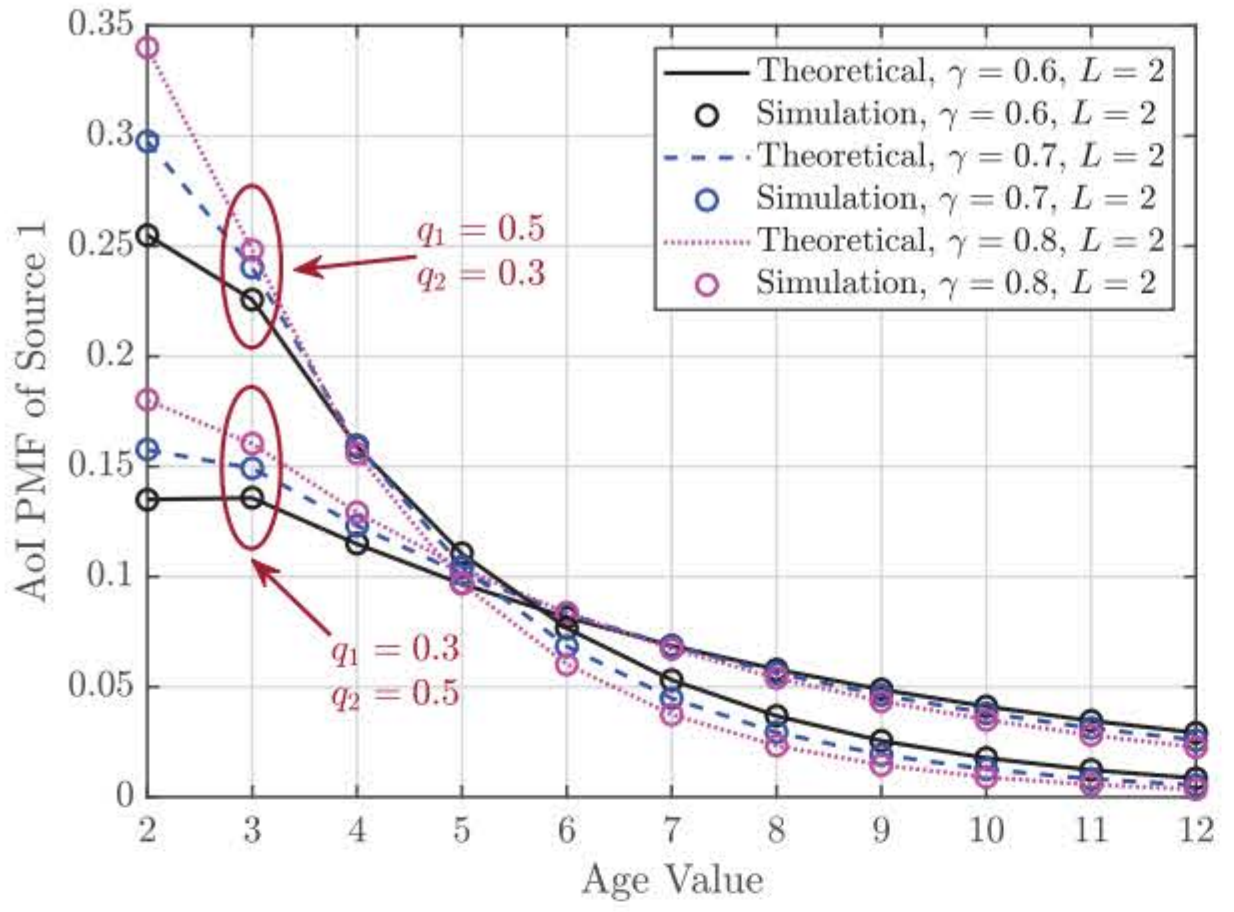

(a) PMF of AoI.

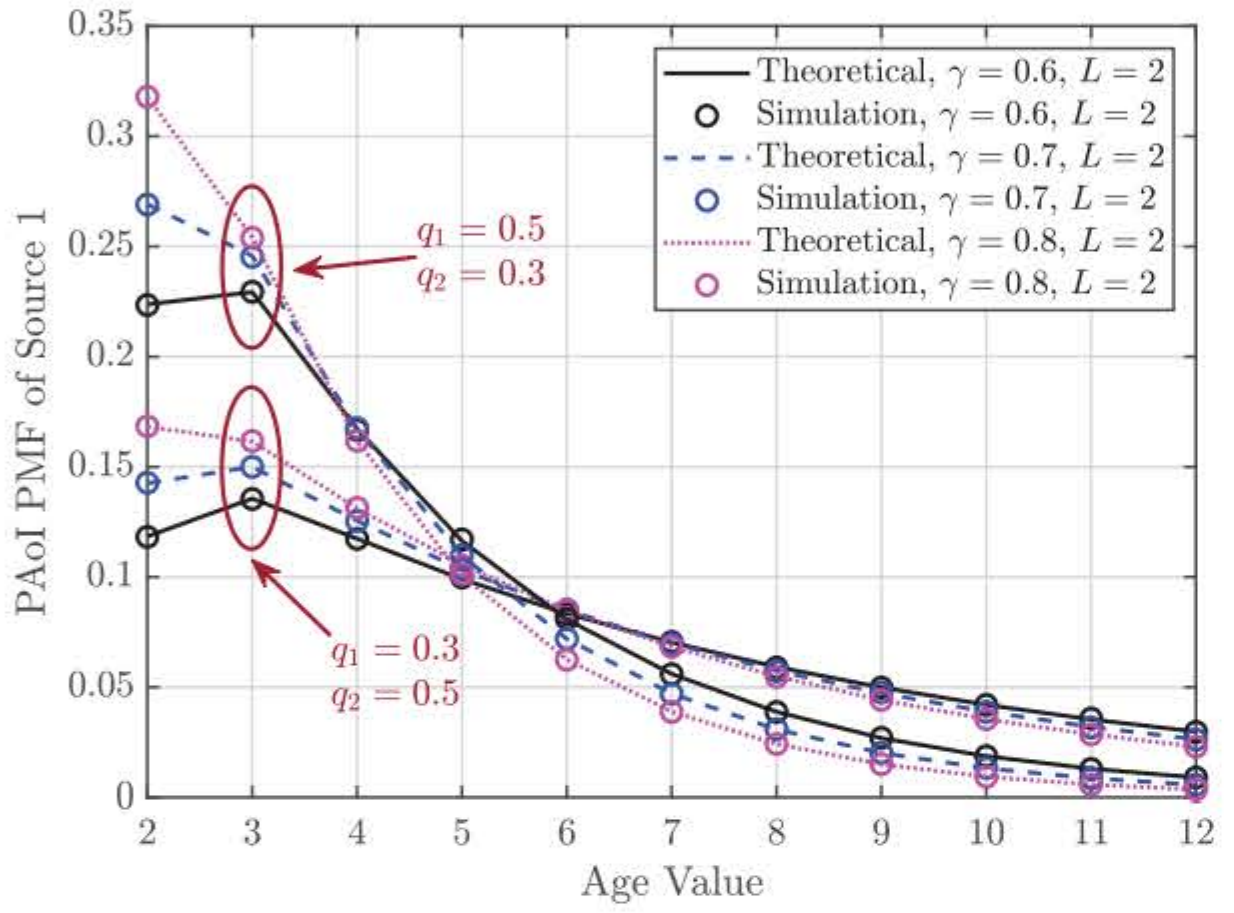

(b) PMF of PAoI.

Fig. 3. PMFs of AoI and PAoI under SATARQ. $N = 2$; $\gamma_i = \gamma$; $L_i = L$.

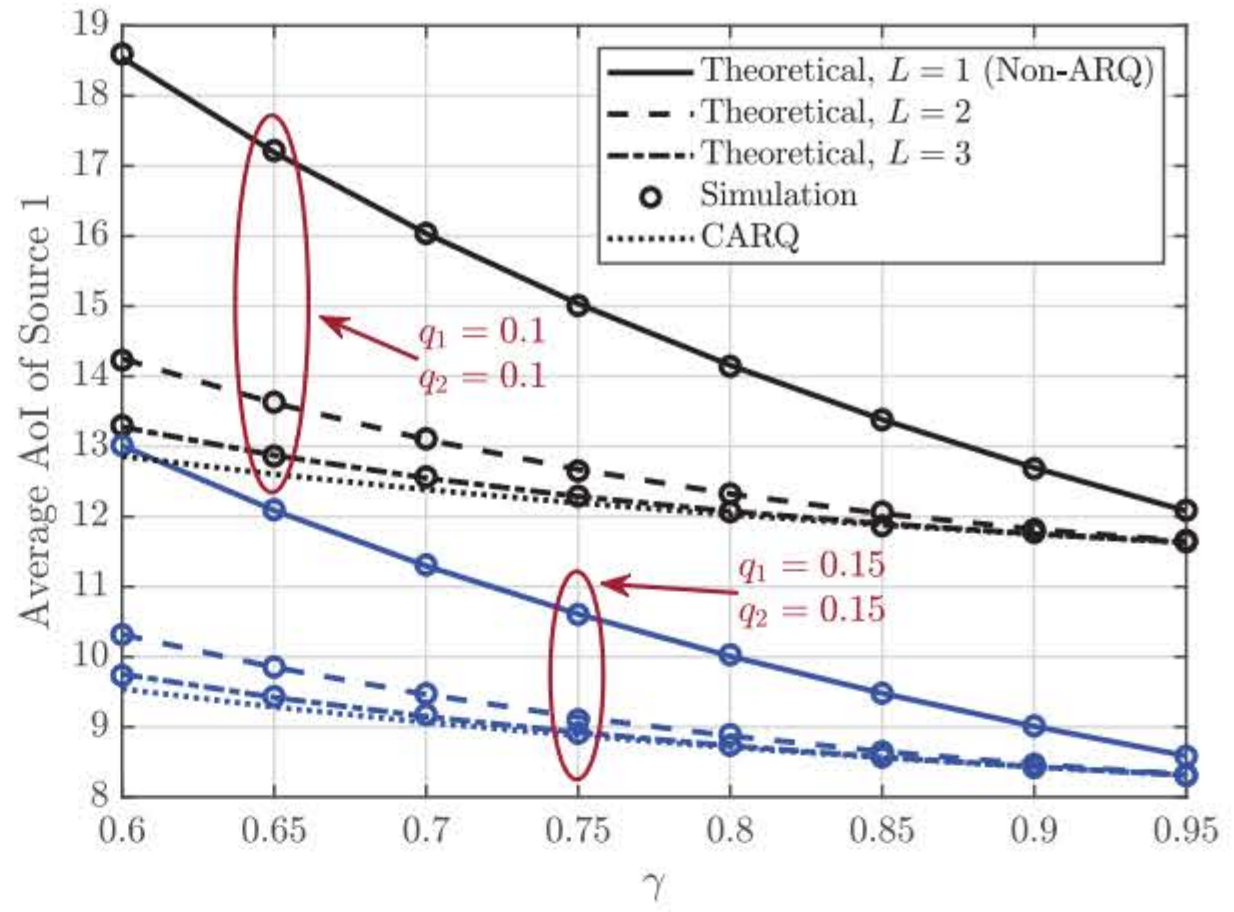

(a) Average AoI.

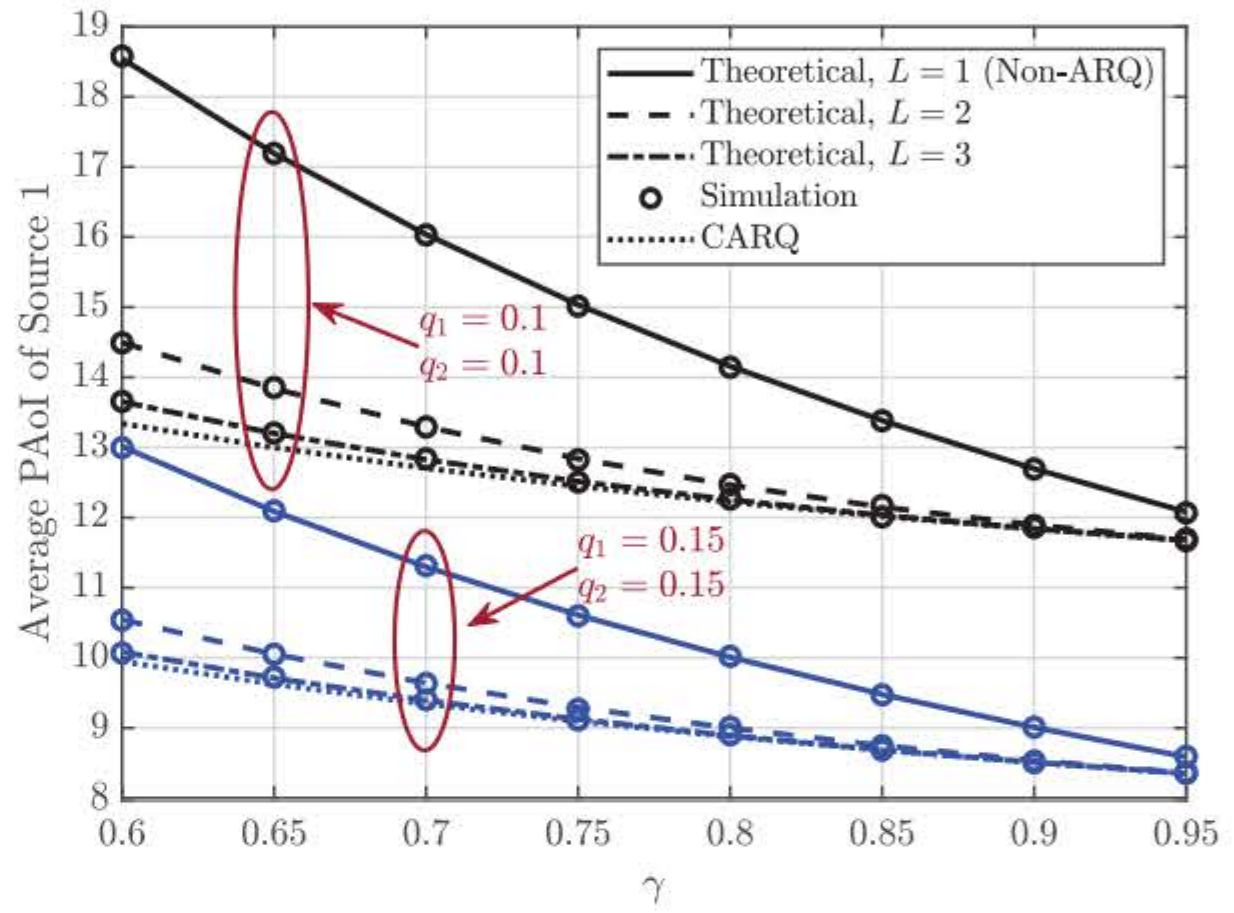

(b) Average PAoI.

Fig. 4. Averages AoI and PAoI under SATARQ. $N = 2$; $\gamma_i = \gamma$; $L_i = L$.

## V. Numerical Results

This section first numerically studies the statistical characteristics of AoI and PAoI. Then, the effectiveness of SATARQ as well as optimizing the MTTs, UGPs, and TPs in enhancing the overall T-E tradeoff is demonstrated. Also, that in improving the overall EE is elaborated. Besides, the limitation of the overall EE for measuring the overall T-E tradeoff is illustrated.

Notably, the models of channel and PTSP used in numerical results are the same as those in Proposition 2, i.e., $\gamma_i = e^{-\frac{k_i}{P_i}}$.

Fig. 3 and Fig. 4 respectively present the PMFs and averages of AoI and PAoI under the SATARQ. The simulations accord with the theoretical results quite well, validating the accuracy of theoretical analysis. It is presented that with the increase of PTSP, *i)* the convergence speeds of the AoI and PAoI PMFs rise, i.e., the AoI and PAoI are more likely to take low values; as well as *ii)* the average AoI and PAoI decrease. These align with the essential relationship between information timeliness and transmission reliability: the more reliable the transmission, the higher the timeliness. It is also found that with the increase of the MTT, the average AoI and PAoI can decrease greatly, especially for low PTSPs, and rapidly converge to that under the CARQ, particularly for large PTSPs. In the case of $q_i = 0.1$ and $\gamma_i = 0.8$, the average AoI is reduced by $12.96\%$ under $L_i = 2$, compared with the case of $L_i = 1$, i.e., the NARQ; In the case of $q_i = 0.1$ and $\gamma_i = 0.95$, the AoI under $L_i = 2$ is $0.15\%$ larger than that under the infinite MTTs, i.e., the CARQ. These indicate that *i)* the SATARQ can significantly outperform the NARQ in the information timeliness, even when the MTT is low; *ii)* the massive retransmissions can be redundant, and the SATARQ with a small MTT can achieve a similar timeliness level as that of the CARQ. Accordingly, the SATARQ is an effective scheme to empower a great timeliness performance.

Figs. 5–7 respectively present the WS of the source-average average AoI and average IoTD power under SATARQ versus

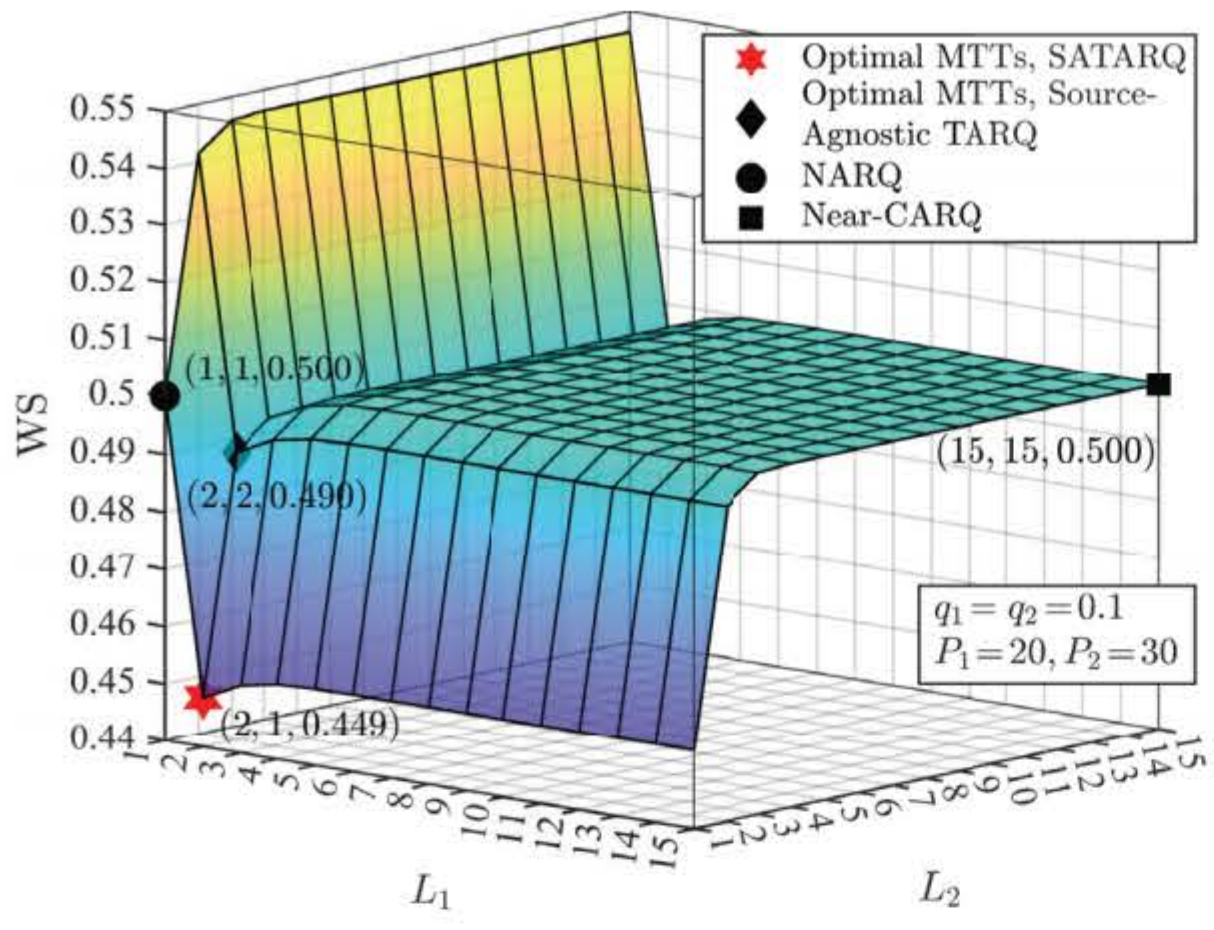

Fig. 5. WS under SATARQ versus MTTs. $N = 2$; $(R_1, R_2) = (2, 1.5)$.

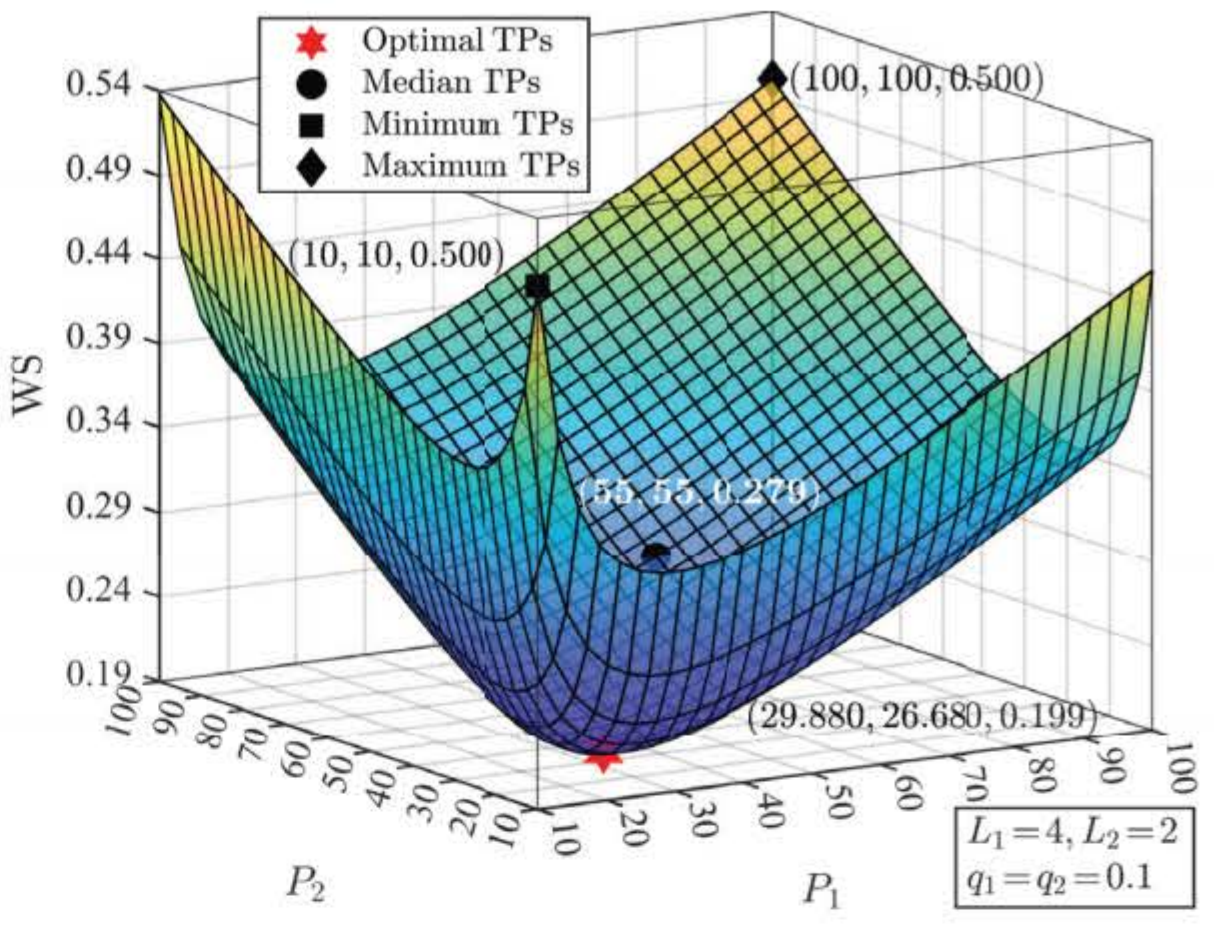

Fig. 7. WS under SATARQ versus TPs. $N = 2$; $(R_1, R_2) = (2, 1.5)$.

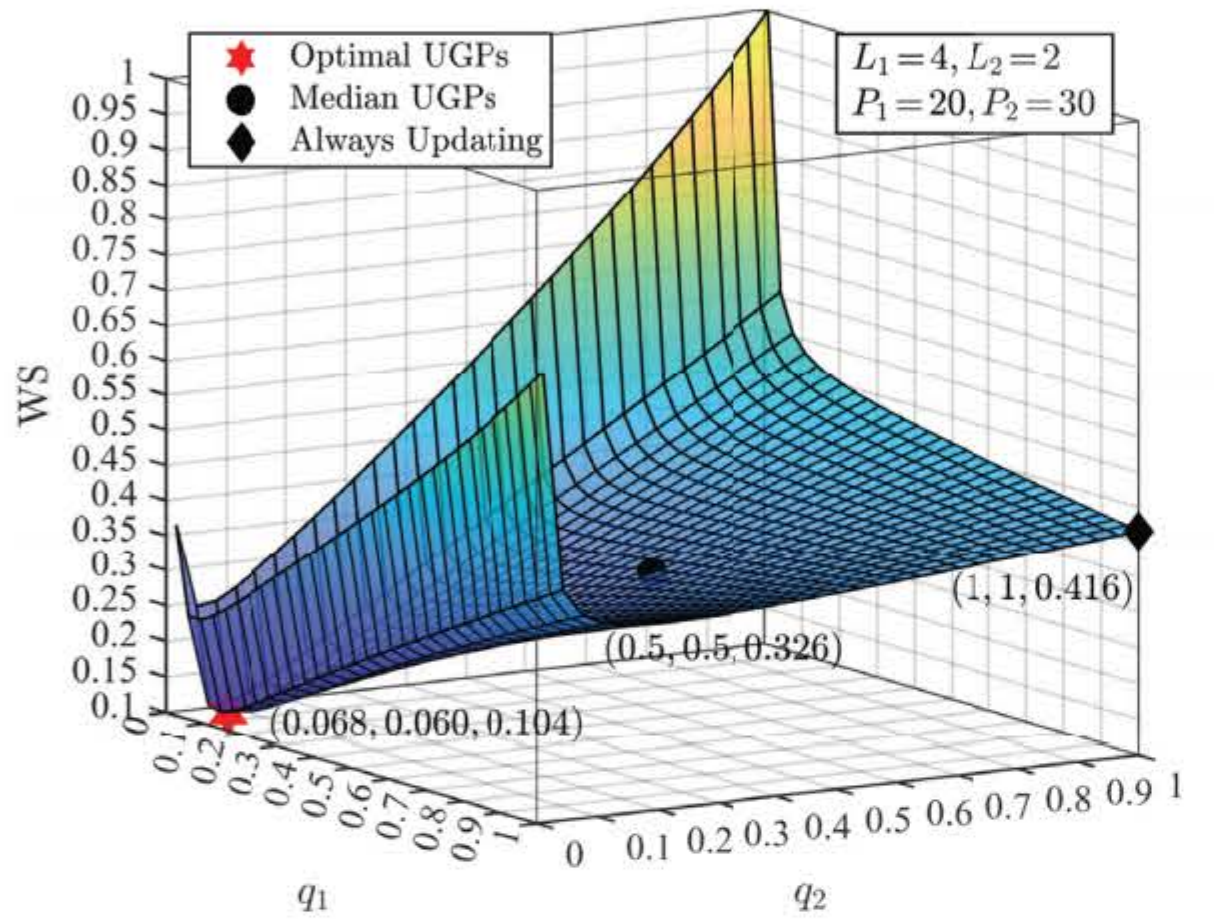

Fig. 6. WS under SATARQ versus UGPs. $N = 2$; $(R_1, R_2) = (2, 1.5)$.

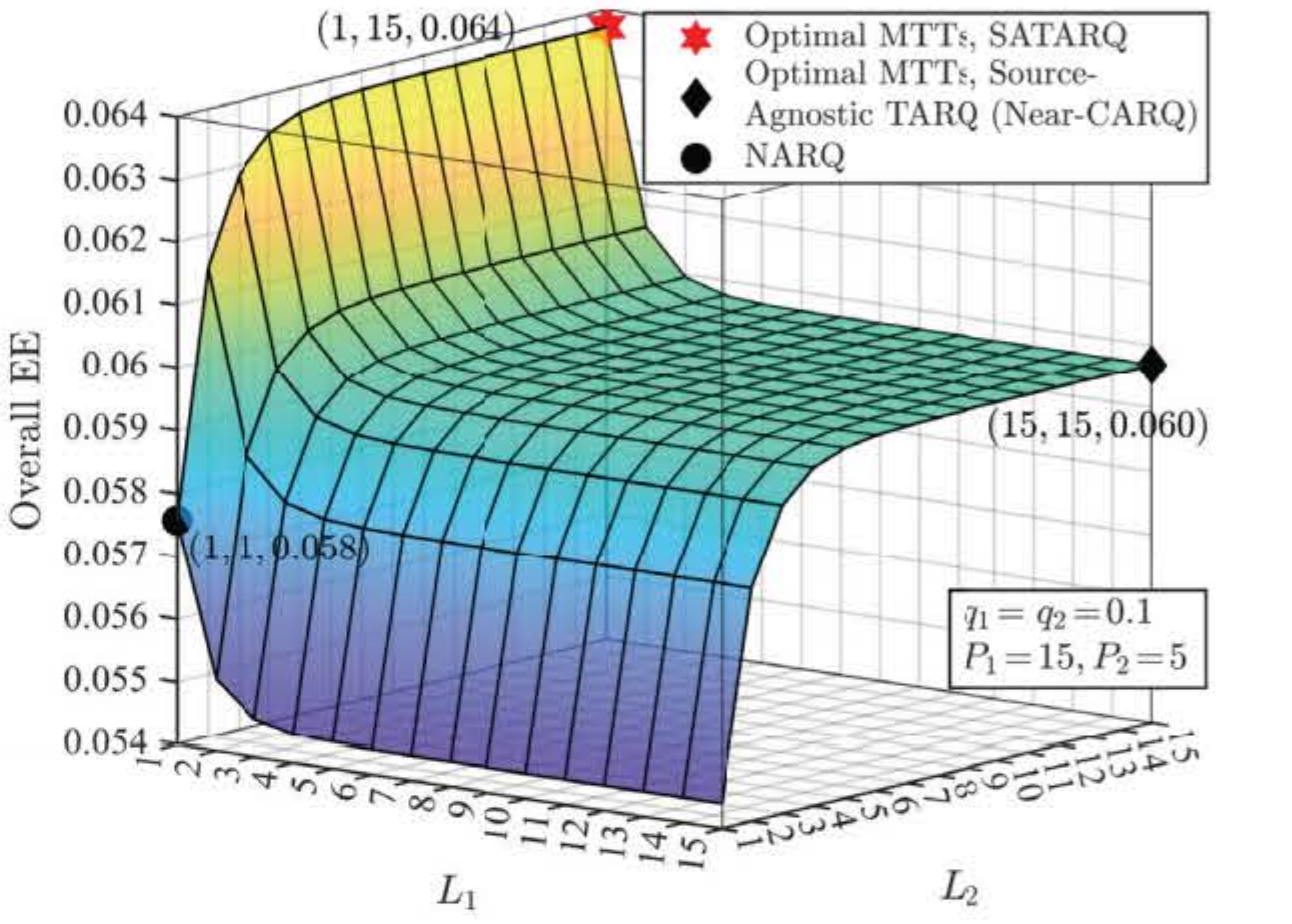

Fig. 8. Overall EE under SATARQ versus MTTs. $N = 2$; $(R_1, R_2) = (2, 1.5)$.

the MTTs, UGPs, and TPs. In particular, the two averages are respectively normalized on the regarded variable ranges. In Figs. 5, 6, and 7, the ranges of the corresponding variables are $1 \leq L_1, L_2 \leq 15$, $0.01 \leq q_1, q_2 \leq 1$, and $10 \leq P_1, P_2 \leq 100$, respectively. The weighting factor is $0.5$. Figs. 5–7 also show the WSs under the SATARQs with the optimal MTTs, optimal UGPs, and optimal TPs, as well as those under some typical schemes. It can be found that the WS varies greatly w.r.t. the MTTs, UGPs, and TPs. Moreover, the WS can be significantly reduced by employing the SATARQ as well as optimizing the MTTs, UGPs, and TPs. For the cases presented in Figs. 5–7, *i)* the WS is reduced by $8.5\%$, $10.3\%$, $10.3\%$ through the SATARQ and the MTT optimization, compared with the source-agnostic TARQ with optimal MTTs, NARQ, and near-CARQ schemes, respectively; *ii)* the WS under the SATARQ is respectively reduced by $68.0\%$ and $74.9\%$ under the optimal UGPs, compared with the schemes of the median UGPs and always updating; as well as *iii)* the TP optimization reduces the WS by $28.5\%$, $60.1\%$, and $60.1\%$, compared with the median, minimum, and maximum TPs, respectively. These indicate the necessity and effectiveness of the SATARQ scheme as well as the optimizations on the MTTs, UGPs, and TPs, in improving the overall T-E tradeoff for multi-source status updating.

Figs. 8–10 respectively present the overall EE under the SATARQ versus the MTTs, UGPs, and TPs. Also, they show the cases of the optimal MTTs, UGPs, and TPs for maximizing the overall EE, as well as other schemes. Similar to the WS, the overall EE varies greatly w.r.t. the MTTs, UGPs, and TPs, and can be remarkably enhanced by adopting the SATARQ and optimizing them. Herein, the minimum and maximum of TPs are considered as $0$ and $15$. For the presented cases, *i)* the overall EE is increased by $6.8\%$ and $10.7\%$ via the SATARQ and the optimization on the MTTs, compared with the optimal source-agnostic TARQ and near-CARQ (the performances of near-CARQ and optimal source-agnostic TARQ are the same in this case); *ii)* the overall EE is raised by $72.3\%$ and $73.1\%$ by optimizing the UGPs, compared with the schemes of the medium UGPs and always updating; as well as *iii)* the overall EE is increased by $54.8\%$ and $121.0\%$ by optimizing the TPs, compared with schemes of medium TPs and maximum TPs.

Moreover, it is presented that *i)* in Fig. 8, the optimal MTT of source $1$ is $1$, which is the minimum value of MTT under the SATARQ scheme; *ii)* in Fig. 9, the optimal UGP of source $1$ is $0$; as well as *iii)* in Fig. 10, the optimal TP of source $1$ is $0$. Note that in the shown cases, the source-specific EE of source $1$ or its maximum is lower than that of source $2$. (In Figs. 8 and

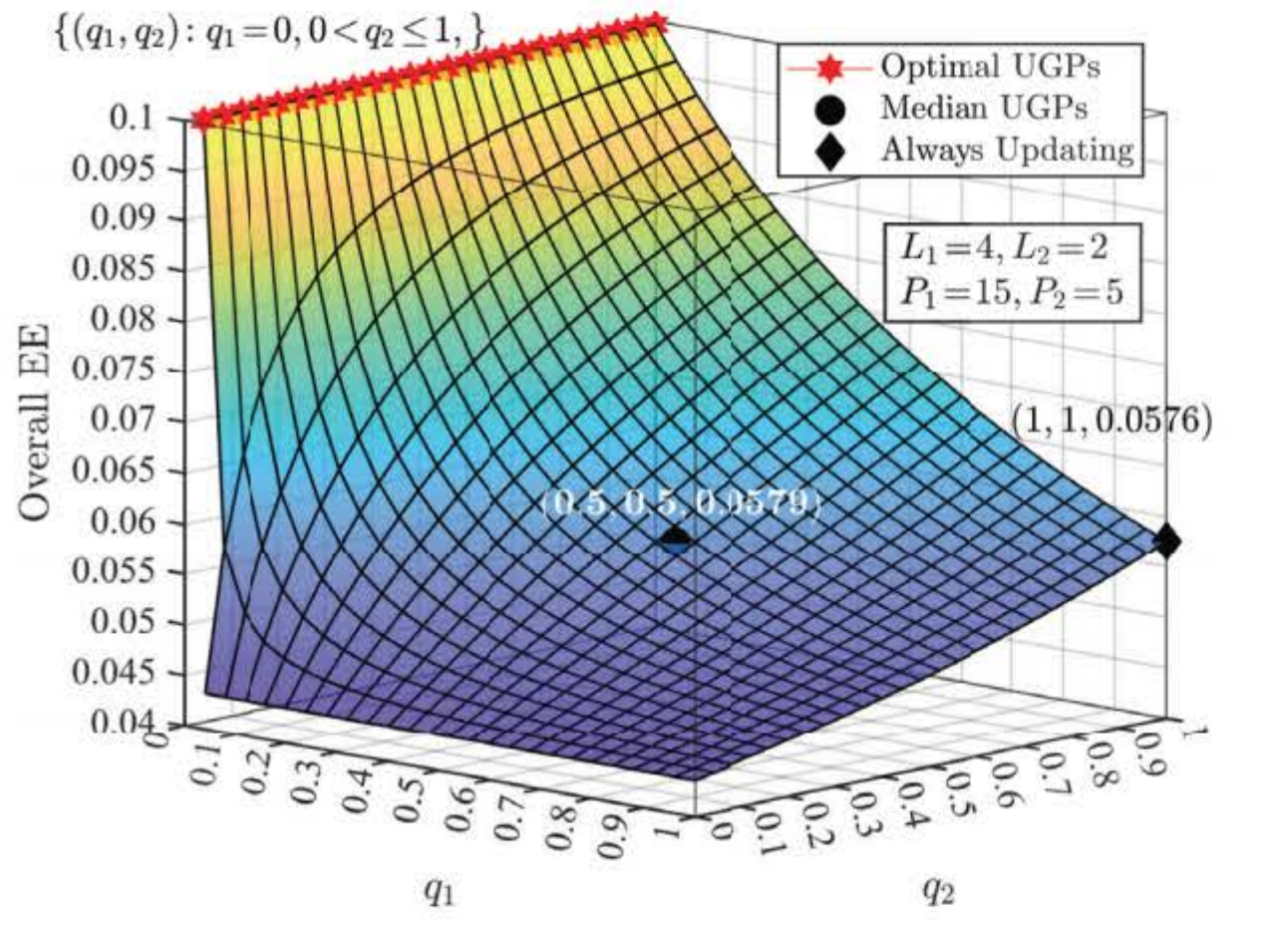

Fig. 9. Overall EE under SATARQ versus UGPs. $N=2$; $(R_1, R_2)=(2, 1.5)$.

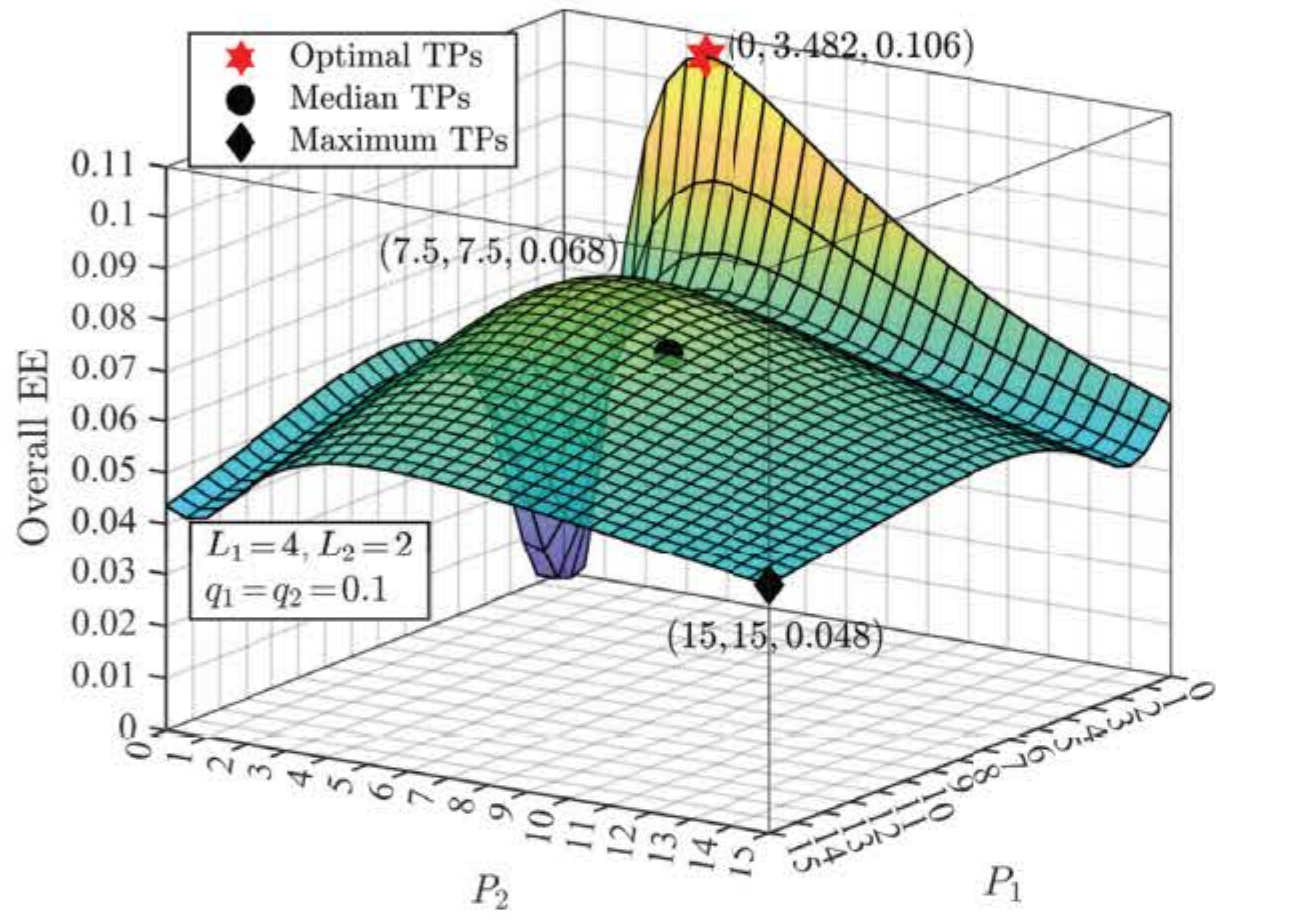

Fig. 10. Overall EE under SATARQ versus TPs. $N=2$; $(R_1, R_2)=(2, 1.5)$.

9, the EEs of sources 1 and 2 are respectively $0.044$ and $0.100$. In Fig. 10, the maximum EEs of sources 1 and 2 are $0.058$ and $0.106$ respectively.) These can reveal that to enhance the overall EE, it could tend to not generate or deliver the updates of the source with low source-specific EE. However, in terms of the information timeliness and system operation, it can be unexpected if the updates of a source cannot be generated or transmitted, which results in the timeliness disaster of this source and even the possible operation breakdown of the whole system. This is a non-negligible limitation of employing the overall EE as the characterization of overall T-E tradeoff. The essential reasons could include that *i)* unlike the arithmetic average, the harmonic average of source-specific average AoIs in the overall EE could not severely penalise the infinite average AoIs, while there is at least one average AoI of finite value; and *ii)* the value of the harmonic average is dominated by the values of the low source-specific average AoIs[2].

[2]The harmonic average can be expressed as $\sum_{i=1}^{N}\frac{\frac{1}{\overline{\Delta}_i-1}}{\sum_{k=1}^{N}\frac{1}{\overline{\Delta}_k-1}}(\overline{\Delta}_i-1)$. That is, it can be regarded as the weighted sum of $\{\overline{\Delta}_i-1\}_{i=1}^{N}$ with weighting factors $\left\{\frac{\frac{1}{\overline{\Delta}_i-1}}{\sum_{k=1}^{N}\frac{1}{\overline{\Delta}_k-1}}\right\}_{i=1}^{N}$. The large AoIs are assigned low weights.

## VI. Conclusion

This work investigated the information timeliness in multi-source wireless status updating. Based on the MDAP method, the AoI and PAoI statistical characteristics under the SATARQ, as well as the average power of the IoTD, were derived. Furthermore, the T-E tradeoff was analyzed from the perspectives of the MTT, UGP, and TP, as well as the EE and its relationship with the timeliness and EC were studied. The numerical results validated the theoretical analysis. Finally, it was demonstrated that the proposed SATARQ scheme as well as the optimization on the MTTs, UGPs, and wireless TPs can effectively improve the overall T-E tradeoff and the overall EE for all sources.